\documentclass[onecolumn,showpacs,preprintnumbers,amsmath,amssymb,nofootinbib]{revtex4-1}
\usepackage{graphicx}
\usepackage{color}

\usepackage[normalem]{ulem}

\renewcommand\sout{\bgroup \color{red}\ULdepth=-.5ex \ULset}
\newcommand\soutb{\bgroup \color{blue} \ULdepth=-.5ex \ULset}

\newcommand{\pslash}{{p\hspace{-5pt}/}}
\newcommand{\kslash}{{k\hspace{-6pt}/}}
\newcommand{\qslash}{{q\hspace{-5pt}/}}
\newcommand{\epslash}{{{\varepsilon}\hspace{-5pt}/}}

\usepackage{dcolumn}
\usepackage{bm}

\usepackage{amsmath}
\begin{document}

\preprint{}

\title{
Quark level and hadronic contributions to the electric dipole moment of charged leptons in the standard model
}

\author{Yasuhiro~Yamaguchi$^{1,2}$}
  \affiliation{$^1$Advanced  Science  Research  Center,  Japan  Atomic  Energy  Agency  (JAEA),  Tokai  319-1195,  Japan}
  \affiliation{$^2$RIKEN Nishina Center, RIKEN, 
  Wako, Saitama 351-0198, Japan}
  \email{yamaguchi.yasuhiro@jaea.go.jp}
\author{Nodoka~Yamanaka$^{3,4}$}
  \affiliation{$^3$Amherst Center for Fundamental Interactions, Department of Physics, University of Massachusetts Amherst, 
Massachusetts
01003, USA}
  \affiliation{$^4$Yukawa Institute for Theoretical Physics, Kyoto University, 
  Kitashirakawa-Oiwake, Kyoto 606-8502, Japan}
  \email{nyamanaka@umass.edu}

\date{\today}

\begin{abstract}
We evaluate the electric dipole moment 
of charged leptons in the standard model, where the complex phase of the Cabibbo-Kobayashi-Maskawa matrix is the only source of CP violation.
We first prove that, at the quark-gluon level, it is suppressed by a factor of $m_b^2 m_c^2 m_s^2$ at all orders of perturbation due to the 
Glashow-Iliopoulos-Maiani mechanism.
We then calculate the hadronic long distance contribution generated by vector mesons at one-loop level.
The $|\Delta S|=1$ weak hadronic interaction is derived using the factorization, and the strong interaction is modeled by the hidden local symmetry framework.
We find that the 
electric dipole moments 
of charged leptons obtained from this hadronic mechanism are much larger than the perturbative four-loop level quark-gluon process, by several orders of magnitude.
\end{abstract}


\maketitle

\section{\label{sec:introduction}Introduction}

The electric dipole moment (EDM) \cite{He:1990qa,Bernreuther:1990jx,Barr:1992dq,Jungmann:2013sga,NaviliatCuncic:2012zza,Khriplovich:1997ga,Ginges:2003qt,Pospelov:2005pr,Raidal:2008jk,Fukuyama:2012np,Engel:2013lsa,Yamanaka:2014,Roberts:2014bka,deVries:2015gea,Yamanaka:2016umw,Yamanaka:2017mef,Chang:2017wpl,Chupp:2017rkp,Safronova:2017xyt,Orzel:2018cui,Yamanaka:2019ifh}
is a very sensitive observable for the detection of the CP violation in many candidate models of new physics beyond the standard model (SM) such as the supersymmetry \cite{Ellis:1982tk,delAguila:1983dfr,Dugan:1984qf,Kizukuri:1992nj,Fischler:1992ha,Ibrahim:1997gj,Ibrahim:1998je,Chang:1998uc,Pokorski:1999hz,Lebedev:2002ne,Chang:2002ex,Pilaftsis:2002fe,Demir:2003js,ArkaniHamed:2004yi,Ellis:2008zy,Lee:2012wa,Yamanaka:2012hm,Yamanaka:2012ia,Yamanaka:2012ep,Dhuria:2014fba,Yamanaka:2014nba,Zhao:2014vga,Li:2015yla,Bian:2016zba,Nakai:2016atk,Cesarotti:2018huy,Zheng:2019knr,Yang:2019aao,Yang:2020ebs}, extended Higgs model \cite{Weinberg:1990me,Barr:1990vd,Leigh:1990kf,Kao:1992jv,Hayashi:1994ha,Barger:1996jc,BowserChao:1997bb,Jung:2013hka,Abe:2013qla,Inoue:2014nva,Bian:2014zka,Chen:2017com,Fontes:2017zfn,Alves:2018kjr,Egana-Ugrinovic:2018fpy,Panico:2018hal,Brod:2018pli,Brod:2018lbf,Okada:2018yrn,Cirigliano:2019vfc,Cheung:2019bkw,Oshimo:2019ylq,Chun:2019oix,Oredsson:2019mni,Fuyuto:2019svr,Eeg:2019eei,Fuchs:2020uoc,Cheung:2020ugr,Kanemura:2020ibp}, Majorana fermion \cite{Ng:1995cs,Archambault:2004td,Fukuyama:2019jiq,Chang:2017vzi}, 
and other interesting models~\cite{Appelquist:2004mn,Fuyuto:2018scm,Dekens:2018bci,Panico:2018hal,Abe:2019wku,Okawa:2019arp,Altmannshofer:2020ywf,Kirpichnikov:2020tcf,Pan:2020qqd,Kirpichnikov:2020lws}.
Among systems in which the EDM may be measured, the charged leptons are the most frequently studied experimentally.
The electron EDM is known to be enhanced by relativistic effect of heavy atomic and molecular systems \cite{Carrico:1968zz,sandars1,sandars2,Flambaum:1976vg,Sandars:1975zz,Labzovskii,Sushkovmolecule,kelly,Kozlov:1994zz,Kozlov:1995xz,flambaumfr1,nayak1,nayak3,nayak2,Natarajrubidium,Mukherjeefrancium,Dzuba:2009mw,Nataraj:2010vn,flambaumybftho,Porsev:2012zx,Roberts:2013zra,Chubukov:2014rba,abe,sunaga,Radziute:2015apa,Denis,Skripnikov,Sunaga:2018lja,Sunaga:2018pjn,Sunaga:2019pfo,Malika:2019jhn,Fazil:2019esp,Talukdar:2020ban}, and it is currently the object of a massive experimental competition \cite{Sandars:1964zz,Weisskopf:1968zz,Stein:1969zz,Player:1970zz,Murthy:1989zz,Abdullah:1990nh,Commins:1994gv,Chin:2001zz,Regan:2002ta,Hudson:2011zz,Sakemi:2011zz,Kara:2012ay,Baron:2013eja,Cairncross:2017fip,Kozyryev:2017cwq,Andreev:2018ayy,Andreev:2018ayy}.
The EDM of the muon is directly measureable in experiments using storage rings \cite{Bennett:2008dy}.
That of the $\tau$ lepton can be extracted by analyzing collider experimental data \cite{Chen:2018cxt,Koksal:2018env, Koksal:2018xyi,Koksal:2018vtt,Dyndal:2020yen}.

In the SM, the Cabibbo-Kobayashi-Maskawa (CKM) matrix \cite{Kobayashi:1973fv} has a CP violating complex phase, so it may generate the EDM.
In the search for new physics beyond the SM, this contribution must be assessed as the leading background.
It is known that, in most cases, it is unobservably much smaller than the experimental sensitivity \cite{Shabalin:1978rs,Shabalin:1980tf,Eeg:1982qm,Eeg:1983mt,Khriplovich:1985jr,Czarnecki:1997bu,Pospelov:1991zt,Booth:1993af,Pospelov:2013sca,Pospelov:1994uf,Khriplovich:1981ca,McKellar:1987tf,Seng:2014lea,Yamanaka:2015ncb,Yamanaka:2016fjj,Lee:2018flm}.
However, the hadronic contribution to the EDM of charged leptons has never been evaluated in the past.
This is just the aim of this paper to quantify it.

In this paper, we first prove that the contribution at the quark-gluon level is suppressed by a factor of $m_b^2 m_c^2 m_s^2$ at all orders of perturbation due to the 
Glashow-Iliopoulos-Maiani (GIM) mechanism.
Next, we calculate the hadronic long distance contribution to the charged lepton EDM generated by vector mesons at one-loop level.
The $|\Delta S|=1$ weak hadronic interaction is derived using the factorization, while the strong interaction is given by the hidden local symmetry framework.
Part of the results have been briefly reported in~\cite{Yamaguchi:2020eub}.
A complete report of our study is given in this article.

This paper is organized as follows.
In the next section, we review the quark-gluon level calculation of the CKM contribution to the EDM of charged leptons and prove that it is actually suppressed by factors of quark masses at all orders of perturbation.
We then describe in Sec. \ref{ref:setup} the setup of the evaluation of the hadronic contribution to the EDMs of charged leptons in the hidden local symmetry framework, with the weak interaction derived with the factorization.
In Sec. \ref{sec:analysis}, we show the result of our calculation and analyze the theoretical uncertainty.
The final section gives the summary of this work.

\section{\label{sec:review}Quark level estimation of the EDM of charged leptons and the GIM mechanism}

Let us first review the previous works on the calculation of the short distance (quark-gluon level) effect to the EDM of charged leptons in the SM.
Since we are supposing that the CP violation is generated by the physical complex phase of the CKM matrix, the Feynman diagrams contributing to the lepton EDM must have at least a quark loop, with sufficient flavor changes so as to fulfill the Jarlskog combination \cite{Jarlskog:1985ht}.
The Jarlskog invariant is given by the product of four CKM matrix elements ($J = {\rm Im}[V_{us} V_{td} V^*_{ud} V^*_{ts}] = (3.18\pm 0.15 )  \times 10^{-5}$ \cite{Tanabashi:2018oca}), so the quark loop must have four $W$ boson-quark vertices.
By noting that the $W$ boson must also be connected to the electron, the two-loop level diagram which has only two vertices in the quark loop does not contribute to the EDM due to the cancellation of the complex phase.

\begin{figure}[htb]
\begin{center}
 \includegraphics[scale=0.8]{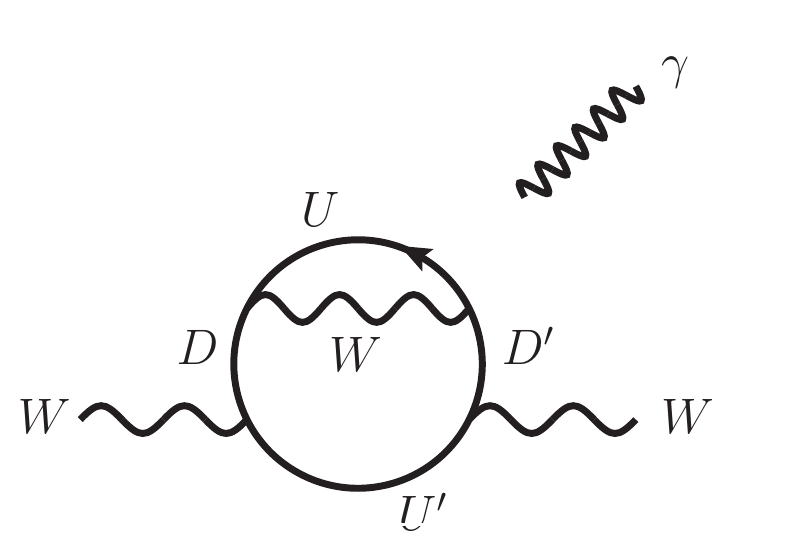}
\caption{
Two-loop level diagram contributing to the EDM of $W$ boson in the SM at the quark level.
The external photon field is attached to all possible propagators.
The sum of all diagrams vanishes, so 
the EDM of charged leptons at the three-loop level, which is generated by attaching the two external $W$ boson lines to the lepton line, also cancels.
}
\label{fig:WEDM}
\end{center}
\end{figure}

The first plausible contribution appears then at the three-loop order~\cite{Hoogeveen:1990cb} (two-loop level diagrams of the EDM of $W$ boson as shown in Fig. \ref{fig:WEDM}, which is attached to the lepton line).
However, extensive three-loop level analyses revealed us that it exactly cancels due to the antisymmetry of the Jarlskog invariant under the flavor exchange (also called the GIM mechanism, a consequence of the CKM unitarity) \cite{Pospelov:1991zt,Booth:1993af,Pospelov:2013sca}.
The cancellation works as follows.
If we can find two quark propagators of the same type 
(up type or down type) 
in the diagram with identical momenta and sandwiched by $W$ boson vertices, the sum of the direct product of these two parts over the $d$-type quark flavors reads
\begin{eqnarray}
&& \hspace{-1em}
\sum_{D\neq D'}
{\rm Im } [V_{U' D} V^*_{U D} V_{U' D'} V^*_{U D'}] 
P_L S_{D} \gamma^\mu P_L \otimes P_L S_{D'} \gamma^\nu P_L
\nonumber\\
&=&
{\rm Im } [V_{U' d} V^*_{U d} V_{U' s} V^*_{U s}] 
( P_L S_{d} \gamma^\mu P_L \otimes P_L S_{s} \gamma^\nu P_L 
-P_L S_{s} \gamma^\mu P_L \otimes P_L S_{d}\gamma^\nu P_L)
\nonumber\\
&&
+{\rm Im } [V_{U' s} V^*_{U s} V_{U' b} V^*_{U b}] 
( P_L S_{s} \gamma^\mu P_L \otimes P_L S_{b} \gamma^\nu P_L 
-P_L S_{b} \gamma^\mu P_L \otimes P_L S_{s}\gamma^\nu P_L)
\nonumber\\
&&
+{\rm Im } [V_{U' b} V^*_{U b} V_{U' d} V^*_{U d}] 
( P_L S_{b} \gamma^\mu P_L \otimes P_L S_{d} \gamma^\nu P_L
-P_L S_{d} \gamma^\mu P_L \otimes P_L S_{b}\gamma^\nu P_L)
\nonumber\\
&=&
{\rm Im } [V_{U' d} V^*_{U d} V_{U' s} V^*_{U s}] 
(P_L k\hspace{-0.45em}/\, \gamma^\mu P_L) \otimes (P_L k\hspace{-0.45em}/\, \gamma^\nu P_L )
\nonumber\\
&&
\times
\Biggl\{
\frac{ 1 }{k^2-m_s^2} \cdot \frac{1 }{k^2-m_d^2}
-\frac{1 }{k^2-m_d^2} \cdot \frac{1 }{k^2-m_s^2}
+\frac{1 }{k^2-m_b^2} \cdot \frac{1 }{k^2-m_s^2}
-\frac{1 }{k^2-m_s^2} \cdot \frac{1 }{k^2-m_b^2}
\nonumber\\
&& \hspace{2em}
+\frac{1}{k^2-m_d^2} \cdot \frac{1 }{k^2-m_b^2}
-\frac{1 }{k^2-m_b^2} \cdot \frac{1}{k^2-m_d^2}
\Biggr\}
\nonumber\\
&=&
0
, \ \ \ 
\end{eqnarray}
where $S_D \equiv \frac{i(k\hspace{-0.4em}/\,+m_D )}{k^2-m_D^2}$.
The projection $P_L \equiv \frac{1}{2} (1-\gamma_5)$ comes from the $W$ boson-quark vertices.
The mass insertions of $S_D$ cancel since odd number of chirality flips is not allowed when $S_D$ is sandwiched by $W$ boson-quark vertices. 
It turns out that the pair of propagators with the same ($u$- or $d$-) type quarks can always be found in the two-loop level contribution to the EDM of $W$ boson, and consequently in the three-loop level diagrams of the EDM of charged leptons.
The most trivial ones are the symmetric diagrams with two quark propagators of the same type, but there are also diagrams which have nonsymmetric insertions of the external photon.
The latter ones can actually be recast into the symmetric form of quark propagators by using the Ward-Takahashi identity \cite{Pospelov:1991zt,Booth:1993af}.
Similar cancellation also occurs in the case of the quark EDM/chromo-EDM \cite{Shabalin:1978rs,Shabalin:1980tf,Eeg:1982qm,Eeg:1983mt,Khriplovich:1985jr,Czarnecki:1997bu} or the Weinberg operator (gluon chromo-EDM) \cite{Pospelov:1994uf}.

\begin{figure}[htb]
 \begin{center}
\includegraphics[scale=0.8]{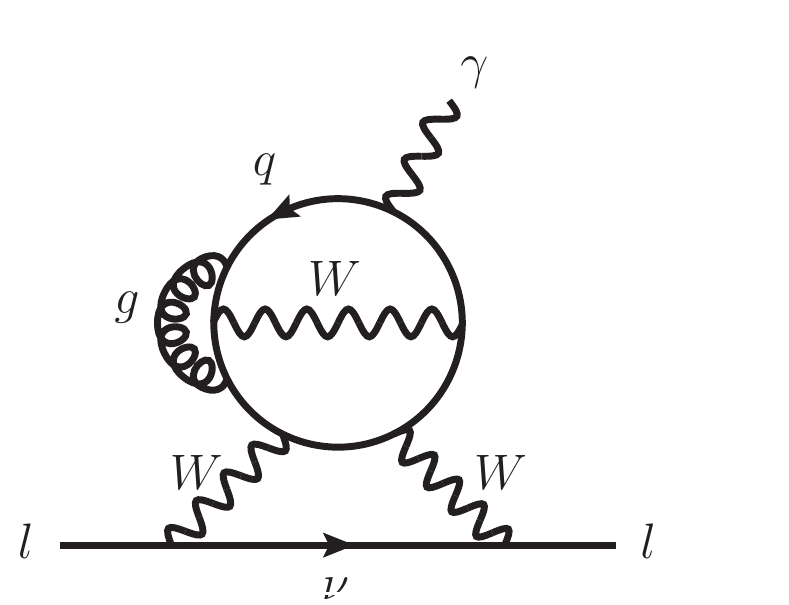}
\caption{
Example of four-loop level diagram contributing to the lepton EDM in the SM at the quark level.
}
\label{fig:electron_EDM_SM_elementary}
 \end{center}
\end{figure}

\begin{figure}[htb]
 \begin{center}
\includegraphics[scale=0.7]{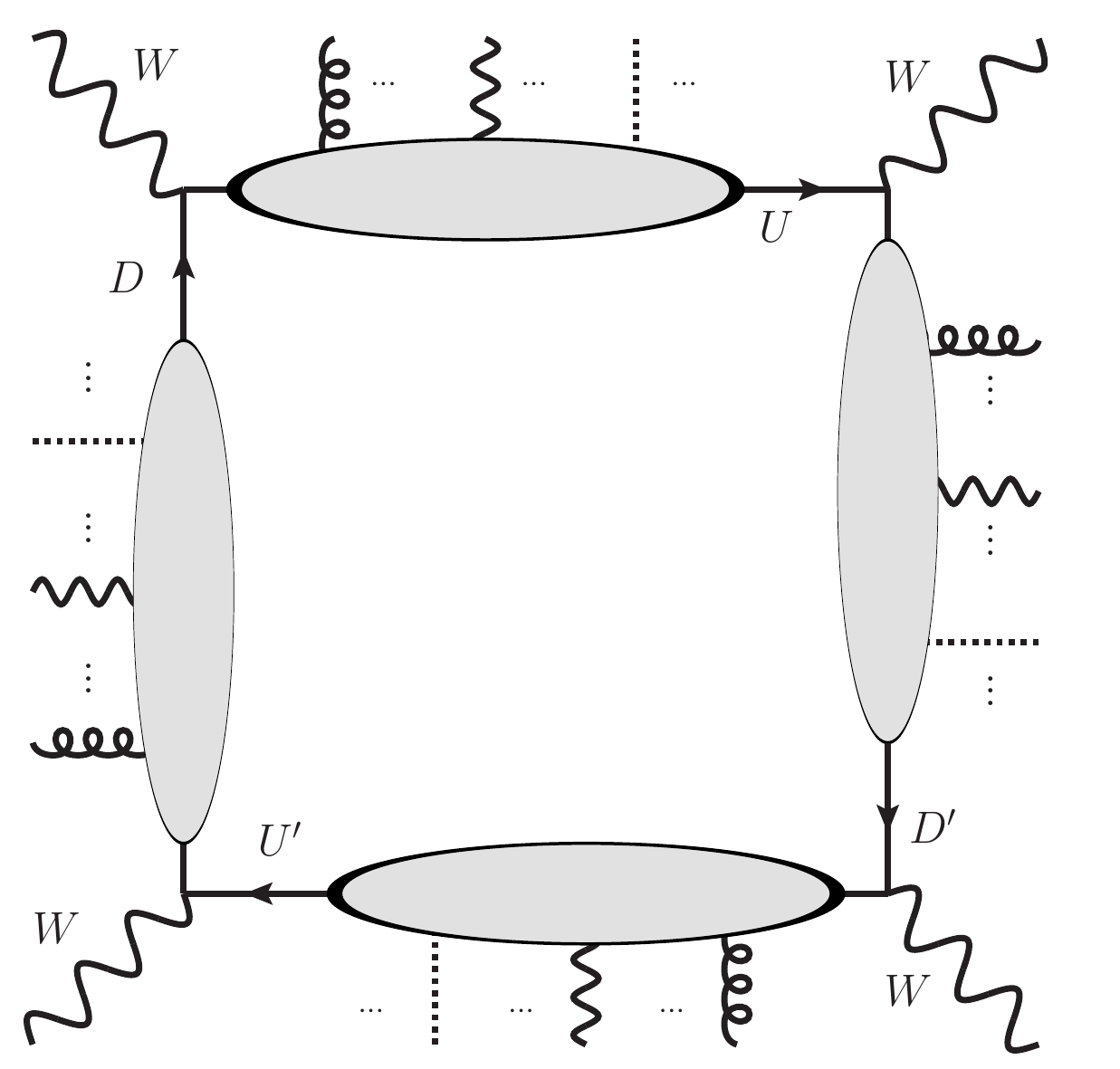}
\caption{
Boson emissions/absorptions of the quark loop with four flavor changing vertices respecting the Jarlskog combination.
The gluon, photon, and the neutral Higgs boson are denoted by the wiggly, wavy, and dashed lines, respectively, and the ellipses means that they each may be of arbitrary number.
The sum of the quark flavors removes the contribution without flip of chirality due to the GIM mechanism.
The emitted bosons have $O(m_W)\approx O(m_t)$ momenta, and they may also form loops, or be connected to other fermion loops, which are not interfering with the flavor structure of the one considered in this figure.
}
\label{fig:GIM_suppression}
\end{center}
\end{figure}

The first nonvanishing contribution which avoids the above symmetric cancellation appears at the four-loop level (see Fig. \ref{fig:electron_EDM_SM_elementary}).
Although the four-loop level contribution has never completely been calculated, it is possible to estimate its size by symmetry consideration.
It is indeed possible to prove that the GIM mechanism \cite{Glashow:1970gm,Ellis:1976fn} always brings additional suppression of quark mass factors $m_q^2$, independently of the order of perturbation.
Let us first consider the quark loop with several insertions of vertices of flavor unchanging (neutral) bosons, i.e. gluons, photons, or Higgs bosons (see Fig. \ref{fig:GIM_suppression}).
We focus on the direct product of the $U$ and $U'$ quark lines with vertex insertions of Fig. \ref{fig:GIM_suppression}, which may be expressed by the Taylor expansion in terms of the quark masses, as follows:
\begin{equation}
\sum_{U \neq U'}
{\rm Im } [V_{U' D} V^*_{U D} V_{U' D'} V^*_{U D'}] 
\sum_{n=0}
a^{(1)}_n m_U^{2n}
\otimes
\sum_{n'=0}
a^{(2)}_{n'} m_{U'}^{2n'}
,
\label{eq:anbn}
\end{equation}
where $a^{(1)}_n$ and $a^{(2)}_{n'}$ are polynomials of the electric charge of up-type quarks, the strong coupling, the inverse of the Higgs vacuum expectation value (appearing from the Yukawa coupling of the Higgs boson after factoring out quark masses), and all momenta carried by the bosons attached to $U$ and $U'$, respectively, which depend on the diagram considered.
Here we took the direct product $\otimes$ to show that the above Taylor expansion also works for the case where Dirac matrices are involved.

We can actually prove that the terms involving $a^{(1)}_0$ and $a^{(2)}_0$ always vanish due to the GIM mechanism.
The case of $a^{(1)}_0 \otimes a^{(2)}_0$ is easy to show, since the sum is just proportional to the sum of Jarlskog invariants, which cancel due to the antisymmetry in the flavor exchange.
The remaining possibilities are $\sum_{n=1} a^{(1)}_n m_U^{2n} \otimes a^{(2)}_0$ and $a^{(1)}_0 \otimes \sum_{n'=1} a^{(2)}_{n'} m_{U'}^{2n'}$ which are also not difficult to treat.
For the former case, we have
\begin{eqnarray}
&&
\sum_{U \neq U'}
{\rm Im } [V_{U' D} V^*_{U D} V_{U' D'} V^*_{U D'}] 
\sum_{n=0}
a^{(1)}_n m_U^{2n}
\otimes
a^{(2)}_0
\nonumber\\
&=&
{\rm Im } [V_{t D} V^*_{u D} V_{t D'} V^*_{u D'}] 
\sum_{n=1}
a^{(1)}_n m_u^{2n}
\otimes
a^{(2)}_0 
+{\rm Im } [V_{t D} V^*_{c D} V_{t D'} V^*_{c D'}] 
\sum_{n=1}
a^{(1)}_n m_c^{2n}
\otimes
a^{(2)}_0 
\nonumber\\
&&
+
{\rm Im } [V_{u D} V^*_{c D} V_{u D'} V^*_{c D'}] 
\sum_{n=1}
a^{(1)}_n m_c^{2n}
\otimes
a^{(2)}_0 
+
{\rm Im } [V_{u D} V^*_{t D} V_{u D'} V^*_{t D'}] 
\sum_{n=1}
a^{(1)}_n m_t^{2n}
\otimes
a^{(2)}_0 
\nonumber\\
&&
+
{\rm Im } [V_{c D} V^*_{t D} V_{c D'} V^*_{t D'}] 
\sum_{n=1}
a^{(1)}_n m_t^{2n}
\otimes
a^{(2)}_0 
+
{\rm Im } [V_{c D} V^*_{u D} V_{c D'} V^*_{u D'}] 
\sum_{n=1}
a^{(1)}_n m_u^{2n}
\otimes
a^{(2)}_0 
\nonumber\\
&=&
{\rm Im } [V_{c D} V^*_{u D} V_{c D'} V^*_{u D'}] 
\Biggl[
\sum_{n=1}
a^{(1)}_n m_u^{2n}
-
\sum_{n=1}
a^{(1)}_n m_c^{2n}
+
\sum_{n=1}
a^{(1)}_n m_c^{2n}
-
\sum_{n=1}
a^{(1)}_n m_t^{2n}
+
\sum_{n=1}
a^{(1)}_n m_t^{2n}
-
\sum_{n=1}
a^{(1)}_n m_u^{2n}
\Biggr]
\otimes
a^{(2)}_0
\nonumber\\
&=&
0
.
\label{eq:Taylorzerothcancel}
\end{eqnarray}
We may repeat the same calculation to show the cancellation for the case of $a^{(1)}_0 \otimes \sum_{n'=1} a^{(2)}_{n'} m_{U'}^{2n'}$ as well.
We thus proved that the leading order CP violation of the quark loop is accompanied by two factors of squared mass of two different up-type quarks to all orders of perturbation in QED, QCD, and Higgs corrections.
We may also exactly repeat the above procedure for the down-type quark contribution which is independent of the up-type ones.
The CP violating part of the quark loop is then at least having a suppression factor of $m_t^2 m_b^2 m_c^2 m_s^2$, which of course persists even if some of the neutral or $W$ bosons are contracted each other or with other quark loops.
The appearance of this factor has actually been already discussed in the general case of the CP violation of 
the CKM matrix~\cite{Jarlskog:1985cw}, and it also appeared in the result of the calculation of
the Weinberg operator which is also generated by a quark loop~\cite{Pospelov:1994uf}.

The presence of the suppression due to quark mass factors, i.e. the cancellation of the zeroth order terms of the Taylor expansion of the quark lines with neutral boson insertions, may also more elegantly be shown using the unitarity of the CKM matrix.
At the order of four $W$ boson-quark vertices, the general flavor structure of the quark loop, with the sum over the flavor taken, is expressed by the following trace
\begin{equation}
{\rm Tr} 
[ V^\dagger Q_U^{(1)} V R_D^{(1)} V^\dagger Q_U^{(2)} V R_D^{(2)} ]
,
\label{eq:flavorstructureV4}
\end{equation}
where $V$ is the $3\times 3$ CKM matrix, and $Q_U^{(k)} \equiv \sum_{n_k=1} a^{(k)}_{n_k} m_{U}^{2n_k}$, $R_D^{(l)} \equiv \sum_{n_l=1} b^{(l)}_{n_l} m_{D}^{2n_l}$ ($k,l = 1,2$) are the down-type and up-type quark lines with arbitrary number of neutral boson insertions, respectively.
We note that $Q_U^{(k)}$ and $R_D^{(l)}$ are $3\times 3$ matrices 
that only have diagonal components.
By taking the zeroth order term of $Q_U^{(1)}$, we have 
\begin{eqnarray}
{\rm Tr} 
[ V^\dagger a_0^{(1)} V R_D^{(1)} V^\dagger Q_U^{(2)} V R_D^{(2)} ]
&=&
a_0^{(1)}
{\rm Tr} 
[ V^\dagger  V R_D^{(1)} V^\dagger Q_U^{(2)} V R_D^{(2)} ]
\nonumber\\
&=&
a_0^{(1)}
{\rm Tr} 
[ R_D^{(2)} R_D^{(1)} V^\dagger Q_U^{(2)} V  ]
\, = \,   a_0^{(1)}
\sum_{i,j=1}^3 
(R_D^{(2)})_i (R_D^{(1)})_i | V_{ij}|^2 (Q_U^{(2)} )_j 
.
\end{eqnarray}
Here we used the unitarity of the CKM matrix $V^\dagger V = 1$, the fact that $R_D^{(2)}, R_D^{(1)}$, and $Q_U^{(2)}$ are diagonal, and that $a_0^{(1)}$ is flavor blind, i.e. proportional to the unit matrix.
The above trace is therefore purely real and the zeroth order terms of the Taylor expansion of the quark lines with neutral boson insertions does not contribute to the EDM.
This expression is exactly equivalent with Eq. (\ref{eq:Taylorzerothcancel}), and at this order $O(V^4)$ the imaginary part only survives when the flavors of all quarks are different, to avoid the appearance of the squared absolute values of the CKM matrix elements.

Next, we have to see higher order corrections with $W$ boson-quark vertices which may be treated in a similar manner.
Here again the unitarity of the CKM matrix plays a crucial role.
Let us consider the case with six $W$ boson-quark vertices.
The general flavor structure of this quark loop, with the flavor summed, looks like
\begin{equation}
{\rm Tr} 
[ V^\dagger Q_U^{(1)} V R_D^{(1)} V^\dagger Q_U^{(2)} V R_D^{(2)} V^\dagger Q_U^{(3)} V R_D^{(3)} ]
.
\end{equation}
We now show that the correction at this order ($V^6$) is not larger than that of $O(V^4)$ which has the quark mass factors $m_t^2 m_b^2 m_c^2 m_s^2$.
A potentially large contribution may arise from the zeroth order terms of the Taylor expansion $a^{(k)}_{0}$ and $b^{(l)}_{0}$.
For example, by considering one such insertion,
\begin{eqnarray}
{\rm Tr} 
[ V^\dagger a^{(1)}_{0} V R_D^{(1)} V^\dagger Q_U^{(2)} V R_D^{(2)} V^\dagger Q_U^{(3)} V R_D^{(3)} ]
&=&
a^{(1)}_{0}
{\rm Tr} 
[ V^\dagger V R_D^{(1)} V^\dagger Q_U^{(2)} V R_D^{(2)} V^\dagger Q_U^{(3)} V R_D^{(3)} ]
\nonumber\\
&=&
a^{(1)}_{0}
{\rm Tr} 
[ R_D^{(3)} R_D^{(1)} V^\dagger Q_U^{(2)} V R_D^{(2)} V^\dagger Q_U^{(3)} V ]
,
\end{eqnarray}
where we again used the unitarity of the CKM matrix.
By noting that $R_D^{(3)} R_D^{(1)}$ is also a diagonal matrix with each component depending only on the mass of one quark flavor, we see that the flavor structure of this contribution is exactly the same as that of the $O(V^4)$ process with neutral boson insertions discussed previously in this section [Fig. \ref{fig:GIM_suppression}, Eqs. (\ref{eq:Taylorzerothcancel}) and (\ref{eq:flavorstructureV4})].
This means that the 
the $O(V^6)$ quark loop having one zeroth order term of the Taylor expansion 
is also having the quark mass factor $m_t^2 m_b^2 m_c^2 m_s^2$.
We also note that the contribution with the three up-type quarks being all top quarks, which may potentially be larger than the $O(V^4)$ terms, has no effect to the EDM, since it will be proportional to three factors of the absolute values of squared CKM matrix elements $|V_{tD}|^2$, i.e. at least a factor of $m_c^2$ or $m_u^2$ is needed.
This analysis may be extended to arbitrary higher orders recursively, since the zeroth order terms $a^{(k)}_{0}$ or $b^{(l)}_{0}$, proportional to the unit matrix, contract two CKM matrix elements $V$ and $V^\dagger$ to form another unit matrix, reducing the flavor trace of $O(V^{2N})$ to $O(V^{2N-2})$.
Since the $O(V^4)$ contribution is having a factor of $m_t^2 m_b^2 m_c^2 m_s^2$, this is also so at $O(V^6)$ and at all other higher orders of $W$ boson-quark vertices ($V$).

We can also show with the above approach the 
cancellation of the quark loop at $O(V^2)$ and at the two-loop level in a more elegant manner.
At $O(V^2)$, we have 
\begin{equation}
{\rm Tr} 
[ V^\dagger Q_U^{(1)} V R_D^{(1)}  ] = \sum_{i=1}^3 H_{ii} (R_D^{(1)})_i
,
\end{equation}
where $H \equiv V^\dagger Q_U^{(1)} V$ is an Hermitian matrix.
Since $R_D^{(1)}$ is diagonal and real, its trace with $H$ is taking only the diagonal elements, which are also real.
There is no room for the imaginary part, so CP is conserved at $O(V^2)$, even accounting for all order corrections of neutral bosons.

At the two-loop level (of the quark loop), we previously saw that we can always find a symmetric set of either up or down-type propagators with the same momentum argument \cite{Pospelov:1991zt,Booth:1993af,Pospelov:2013sca}.
We may then write the trace as
\begin{equation}
{\rm Tr} 
[ V S_{D} V^\dagger Q_U^{(1)} V S_{D} V^\dagger Q_U^{(2)} ] 
= \sum_{i,j=1}^3 H'_{ij} (Q_U^{(1)})_j H'_{ji} (Q_U^{(2)})_i
= \sum_{i,j=1}^3 |H'_{ij}|^2 (Q_U^{(1)})_j (Q_U^{(2)})_i
,
\end{equation}
where we used the 
Hermiticity of $H' \equiv V S_D V^\dagger$.
Due to the absolute value, there is no CP violation, and there is thus no contribution to the EDM of charged leptons at the three-loop level.
We also see that, if the symmetry between the two $S_D$ is destroyed, the two $H'$ will no longer be complex conjugates, and the imaginary part will be generated.

Let us now estimate the EDM of charged leptons according to the above discussion.
The correct dimensional analysis of the four-loop level contribution according to the above proof therefore yields
\begin{equation}
d_l
\sim
\frac{e J \alpha_s 
\alpha_{\rm QED}^3
m_l m_b^2 m_c^2 m_s^2}{\sin^6 \theta_W m_t^8 (4\pi)^4}
,
\end{equation}
which is transcribed to
\begin{eqnarray}
d_e
&=&
O(10^{-50}) e \, {\rm cm}
,
\label{eq:eEDMquark}
\\
d_\mu
&=&
O(10^{-48}) e \, {\rm cm}
,
\label{eq:muEDMquark}
\\
d_\tau
&=&
O(10^{-47}) e \, {\rm cm}
.
\label{eq:tauEDMquark}
\end{eqnarray}
Here we did not consider the logarithmic enhancement which may enlarge the above values by one or two orders of magnitude.
Nevertheless, these results are actually telling us that the short distance contribution is extremely small.
From this analysis, we see that the enormous suppression of the EDM of charged leptons is not due to the fact that it appears at the four-loop level, but rather due to the 
cancellation by the GIM mechanism.

We stress that this suppression mechanism does apply only when all momenta involved are of $O(m_W \sim m_t)$.
In the case where nonperturbative physics is relevant in the infrared region, the coefficients $a_n$, $b_{n'}$ of Eq. (\ref{eq:anbn}) may be enhanced by $O(1/\Lambda_{\rm QCD}^2 ) = O({\rm GeV}^{-2})$ factors.
In the next section, we recast the soft momentum physics into phenomenological hadron physics where the weak interacting hard part is given by low energy constants, which are also calculated with phenomenological models.

\section{\label{ref:setup}Setup of the calculation}

\subsection{The long distance effect}

\begin{figure}[htb]
 \begin{center}
\includegraphics[width=0.4\linewidth]{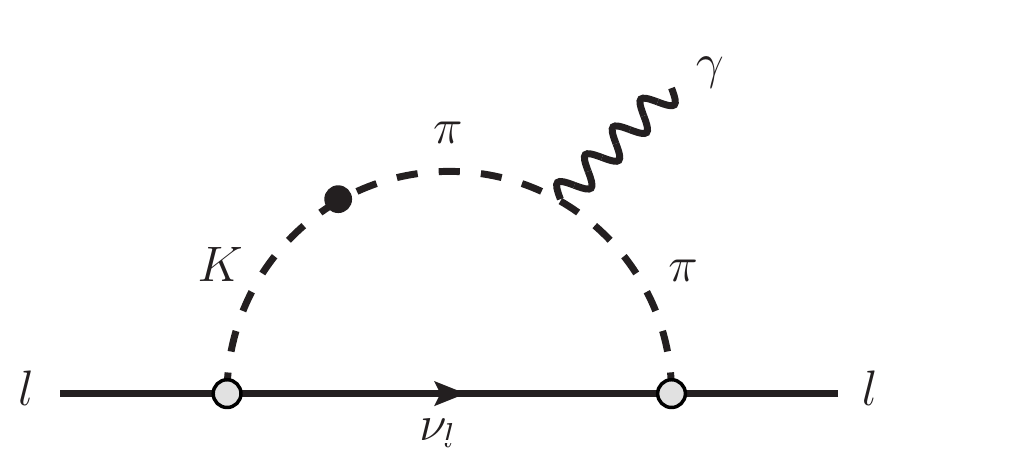}
\caption{Example of a one-loop level contribution to the EDM of the charged
lepton $l$ generated by pseudoscalar mesons ($\pi, K$). 
The grey and black blobs denote the weak interaction.
This 
na\"ively leading diagram is suppressed by the chirality flips of the
pseudoscalar meson-lepton vertices (grey blobs).
}
\label{fig:electron_EDM_pseudoscalar_loop}
 \end{center}
\end{figure}

\begin{figure}[htb]
 \begin{center}
\includegraphics[width=1.0\linewidth]{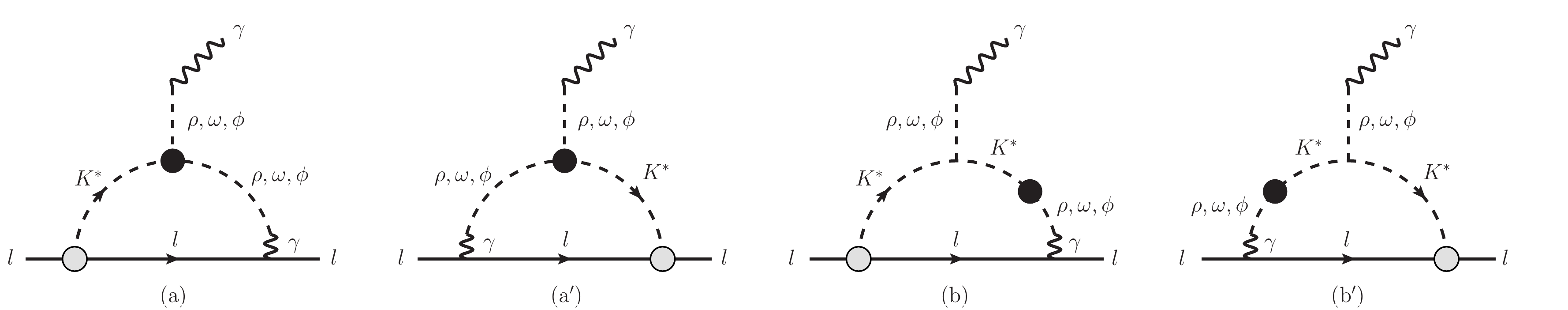}
\caption{
Long distance contribution to the EDM of charged lepton $l$ ($=e,\mu , \tau$) in the SM.
The diagrams $(a)$ and $(a^\prime)$ ($(b)$ and $(b^\prime)$) are the contribution with the weak (strong) three vector meson interactions.
  There are also diagrams with the $\bar{K}^\ast$ propagator, which are not displayed.
The grey blob denotes the $|\Delta S|=1$ 
semileptonic effective interaction, while the black one is the $|\Delta S|=1$ (two- and three-point) vector meson interactions which combine each other to form the Jarlskog invariant.
}
\label{fig:electron_EDM_long-distance}
 \end{center}
\end{figure}

The leading order contribution of the CKM matrix to the lepton EDM is constructed with at least two $W$ boson exchanges. 
To avoid severe GIM cancellation as we saw in the previous section, we have to split the short distance flavor changing process at least into two parts at the hadron level (the long distance effect), while keeping the Jarlskog combination of the CKM matrix elements.
The largest long distance contribution should involve unflavored and $|S|=1$ mesons rather than 
the heavy flavored ($c,b$) ones.
Another important condition is that the charged lepton EDM is generated by one-loop level diagrams involving vector mesons, because the interaction of pseudoscalar mesons with the lepton will change the chirality, suppressing the EDM by at least by a factor of $m_l^2$ ($l=e, \mu , \tau$) 
(for an example of a one-loop level diagram with pseudoscalar mesons
suppressed by chirality flips, see Fig.~\ref{fig:electron_EDM_pseudoscalar_loop}).
The charged lepton EDM is then generated by diagrams involving a $K^*$ meson.
The one-loop level diagrams must not have a neutrino in the intermediate state of the long distance process, since the small neutrino mass will not provide sufficient chirality flip required in the generation of the EDM. 
Moreover, if the process contains two weak $K^*$-charged lepton vertices, the chirality selection will not allow an EDM.
The $K^*$ meson must therefore change to an unflavored meson which in turn becomes a photon which will be absorbed by the charged lepton.
Under such restrictions, 
we may draw diagrams shown in Fig. \ref{fig:electron_EDM_long-distance}.
We note that diagrams with external photons attached to internal lepton propagator cancel when transposed diagrams are summed.

\subsection{Hidden local symmetry}

Let us now give the interactions to calculate the diagrams of Fig. \ref{fig:electron_EDM_long-distance}.
It is convenient to describe the $|\Delta S|=0$ vector meson interactions with the hidden local symmetry (HLS) \cite{Bando:1984ej,Bando:1984pw,Fujiwara:1984mp,Bando:1985rf,Bando:1987br,Meissner:1987ge,Kaiser:1990yf,Harada:1992np,Klingl:1996by,Harada:2000kb,Harada:2003jx}.
The HLS is a framework introduced to extend the domain of applicability of chiral perturbation to include vector meson resonances, and it is successful in phenomenology \cite{Harada:2003jx}.
The effective Lagrangian for three vector mesons is given by
\begin{align}
 {\cal L}_{3V}&=ig{\rm tr}
 \left[\left(\partial_\mu    V_\nu-\partial_\nu   V_\mu\right)     V^\mu    V^\nu\right],
 \label{eq:LVVV}
\end{align}
where the vector meson matrix $V^\mu$ is given by
\begin{align}
 V^\mu&=\left(
 \begin{array}{ccc}
  \frac{\rho^0}{\sqrt{2}}+\frac{\omega}{\sqrt{2}}&\rho^+ &K^{\ast+} \\
  \rho^-&-\frac{\rho^0}{\sqrt{2}}+\frac{\omega}{\sqrt{2}} &K^{\ast 0} \\
  K^{\ast -}& \bar{K}^{\ast 0}&\phi \\
 \end{array}
 \right)^\mu,
\end{align}
where $g=m_\rho/(2f_\pi)$ with the pion decay constant $f_\pi=93$ MeV.

The effective Lagrangian for vector meson and photon 
is given by~\cite{Klingl:1996by}
\begin{align}
 {\cal L}_{\gamma V}&=-\sqrt{2}\frac{em^2_\rho}{g_\gamma}A_\mu {\rm tr}(QV^\mu)
 =-\frac{em^2_\rho}{g_\gamma}A_\mu
 \left(\rho^{0\mu}+\frac{1}{3}\omega^\mu-\frac{\sqrt{2}}{3}\phi^\mu \right), 
\end{align}
where $g_\gamma=5.7$  
and
\begin{align}
 Q&=\left(
 \begin{array}{ccc}
  \frac{2}{3}&0 &0 \\
  0&-\frac{1}{3} &0 \\
  0& 0& -\frac{1}{3}\\
 \end{array}
 \right).
\end{align}

\subsection{$K^*$-lepton interaction}

Let us now model the weak interaction at the hadron level.
From Fig. \ref{fig:electron_EDM_long-distance}, the $|\Delta S|=1$ weak interaction appears in the $K^*$-lepton interaction and in the interacting vertices between $K^*$ and other vector mesons.
Since the neutrino cannot appear in Fig. \ref{fig:electron_EDM_long-distance}, the interaction between $K^*$ and the charged lepton must be at least a one-loop level process at the quark level.
Then the best solution is to attribute the CKM matrix elements $V_{cs}V_{cd}^*$ or $V_{ts}V_{td}^*$ to the $K^*$-lepton interaction, and $V_{ud}V_{us}^*$ to the $K^*$-vector meson interactions.
The latter attribution will maximize the $|\Delta S|=1$ vector meson interactions, since $V_{ud}V_{us}^*$ is given from the tree level $|\Delta S|=1$ four-quark interaction.

\begin{figure}[htb]
 \begin{center} 
\includegraphics[width=15cm]{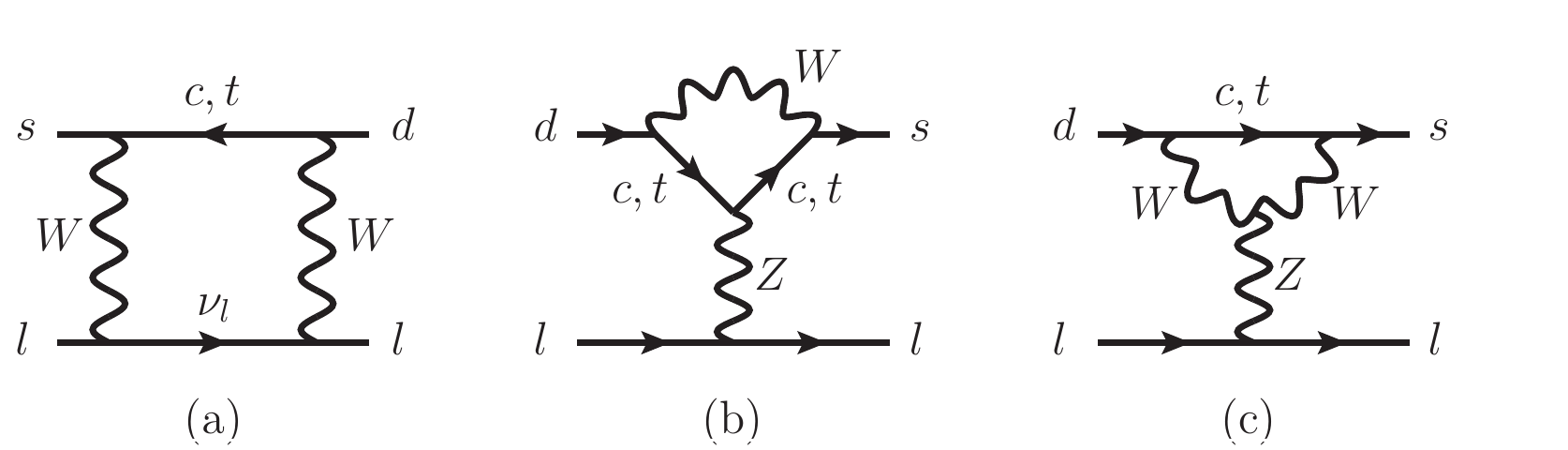}
\caption{
Short distance contribution to the $\Delta S=-1$ 
semileptonic ($K^*$-charged lepton) interaction.
Here we have $l=e,\mu , \tau$.
}
\label{fig:semileptonic_one-loop}
 \end{center}
\end{figure}

The parity violating effective interaction between $K^*$ and the charged lepton is given by
\begin{equation}
{\cal L}_{K^*ll}
=
g_{K^*ll} 
K^{*}_\mu
\bar l \gamma^\mu \gamma_5 l
 + {\rm (H.c.)} 
,
\end{equation}
where $K^*_\mu$ is the field operators of the $K^*$ meson.
In the zero momentum exchange limit, the coupling constant is given by 
\begin{equation}
{\rm Im} ( g_{K^*ll} ) \varepsilon^{K^{*}}_\mu
=
{\rm Im} ( V_{ts}^*V_{td} )
\langle 0 | \bar s \gamma_\mu d | K^* \rangle 
I_{dsll}
, 
\end{equation}
where we fixed the complex phases of $V_{ud}V_{us}^*$ to be real.
The $K^*$ meson matrix element is given by 
\begin{equation}
\langle 0 | \bar s \gamma_\mu d | K^* \rangle = m_{K^*} f_{K^*} \varepsilon^{K^{*}}_\mu 
,
\end{equation}
where $\varepsilon^{K^{*}}_\mu $, $m_{K^*}=890$ MeV and $f_{K^*}= 204$ MeV \cite{Neubert:1997uc,Grossmann:2015lea,Straub:2015ica,Chang:2018aut} are the polarization vector, the mass, and the decay constant of $K^*$, respectively. 
The quark level amplitude $I_{dsll}$ can be obtained by calculating the one-loop level diagrams of Fig. \ref{fig:semileptonic_one-loop}.
By neglecting all external momenta [which are $O(\Lambda_{\rm QCD})$] and imposing $m_t, m_W \gg m_c$, the amplitude of the diagrams of Fig. \ref{fig:semileptonic_one-loop} is given by
\begin{widetext}
\begin{eqnarray}
{\rm Im} ({\cal M}_{\rm (a)}^{K^\ast ll})
&\approx&
\frac{ - \alpha_{\rm QED}^2 {\rm Im} (V_{ts} V_{td}^* )}{
16
\sin^4 \theta_W}
\frac{m_t^2}{m_t^2-m_W^2}
\Biggl\{
\frac{1}{m_W^2}
+\frac{1}{m_t^2-m_W^2} \ln \,
\Biggl(
\frac{m_W^2}{m_t^2}
\Biggr)
\Biggr\}
\bar u_e \, \gamma^\mu \gamma_5 u_e \cdot 
\bar u_d \, \gamma_\mu (1-\gamma_5) u_s
,
\label{eq:semileptonicamplitude1}
\\
{\rm Im} ({\cal M}_{\rm (b)}^{K^\ast ll})
&\approx &
\frac{-\alpha_{\rm QED}^2 {\rm Im} (V_{ts}V_{td}^*)
}{16 \sin^4 \theta_W \cos^2 \theta_W m_Z^2}
\bar u_e \gamma^\mu \gamma_5 u_e \, \bar u_d\gamma_\mu (1-\gamma_5)  u_s
\nonumber\\
&& 
\times 
\frac{m_t^2}{m_t^2-m_W^2}
\left\{
\Biggl( \frac{1}{2} -\frac{2}{3} \sin^2 \theta_W \Biggr) 
\ln \Biggl(\frac{m_W^2}{m_t^2} \Biggr)
-
\Biggl( \frac{1}{2} +\frac{2}{3} \sin^2 \theta_W \Biggr)
\Biggl[
1
+\frac{m_W^2}{m_t^2-m_W^2} \ln \,
\Biggl(
\frac{m_W^2}{m_t^2}
\Biggr)
\Biggr]
\right\}
,
\label{eq:semileptonicamplitude2}
\\
 {\rm Im} ({\cal M}_{\rm (c)}^{K^\ast ll})
&\approx &
\frac{ 
3 \alpha_{\rm QED}^2 {\rm Im} (V_{ts}V_{td}^*)
}{
16
\sin^4 \theta_W m_Z^2}
\frac{m_t^2}{(m_W^2-m_t^2)^2}
\Biggl\{
m_W^2
-m_t^2
\Biggl[
1+\ln \,
\Biggl(
\frac{m_W^2}{m_t^2}
\Biggr)
\Biggr]
\Biggr\}
\,
\bar u_e \gamma^\mu \gamma_5 u_e
\cdot
\bar u_d \, \gamma_\mu (1-\gamma_5) u_s
,
\label{eq:semileptonicamplitude3}
\end{eqnarray}
\end{widetext}
where $\sin^2\theta_W = 0.23122$~\cite{Tanabashi:2018oca}.
The diagram (c) is the largest, but all of them are of the same order.
The numerical value of the total $I_{dsll}$ is
\begin{equation}
I_{dsll}
=
3.2 \times 10^{-8}\, {\rm GeV}^{-2}
,
\end{equation}
which is quite consistent in absolute value with that of the 
na\"ive dimensional analysis 
$I_{dsll} \sim \frac{\alpha_{\rm QED}^2}{4\sin^4 \theta_W m_W^2} \sim 4.3 \times 10^{-8}\, {\rm GeV}^{-2}$.
We note that Eqs. (\ref{eq:semileptonicamplitude1}), (\ref{eq:semileptonicamplitude2}), and (\ref{eq:semileptonicamplitude3}) all contain a factor of $\frac{m_t^2}{m_W^2-m_t^2}$ which is due to the GIM cancellation.
This shows that if we invert the up-type and down-type quarks, the resulting meson-charged lepton couplings will be suppressed by a factor of $m_D^2/m_W^2$ ($D=d,s,b$).

\subsection{
$|\Delta S|=1$ vector meson transition and three-vector meson interaction
}
We now model the $|\Delta S|=1$ vector meson transition and three-vector meson interaction using the factorization.
For that, we have to determine the Wilson coefficients of the quark level $|\Delta S|=1$ processes.
We chose the $|\Delta S|=1$ case because it is the only allowed flavor change at low energy scale.
At the scale just below the $W$ boson mass ($m_W = 80.4$ GeV), we have the following $|\Delta S| =1$ effective 
Hamiltonian
\begin{eqnarray}
{\cal H}_{eff} (\mu = m_W)
&=&
\frac{G_F}{\sqrt{2}}
\Biggl\{
\sum_{i=1,2} C_i (\mu = m_W) [ V_{us}^* V_{ud} Q_i + V_{cs}^* V_{cd} Q_i^c ]
- \sum_{j=3}^6 C_j (\mu = m_W) V_{ts}^* V_{td} Q_j
\Biggr\}
 +{\rm H.c.} 
,\ \ \ \ 
\label{eq:effhamimw}
\end{eqnarray}
with the Fermi constant $G_F = 1.16637 \times 10^{-5} {\rm GeV}^{-2}$ \cite{Tanabashi:2018oca}.
Here $Q^q_1$, $Q^c_1$, $Q^q_2$, $Q^c_2$, and $Q_j$ ($j=3 \sim 6$) are defined as \cite{Buras:1991jm,Buchalla:1995vs}
\begin{eqnarray}
Q_1 
&\equiv &
\bar s_\alpha \gamma^\mu (1-\gamma_5) u_\beta 
\, 
\bar u_\beta \gamma_\mu (1-\gamma_5) d_\alpha
,
\label{eq:q1}
\\
Q_1^c 
&\equiv &
\bar s_\alpha \gamma^\mu (1-\gamma_5) c_\beta 
\, 
\bar c_\beta \gamma_\mu (1-\gamma_5) d_\alpha
,
\label{eq:q1c}
\\
Q_2 
&\equiv &
\bar s_\alpha \gamma^\mu (1-\gamma_5) u_\alpha 
\, 
\bar u_\beta \gamma_\mu (1-\gamma_5) d_\beta
,
\label{eq:q2}
\\
Q_2^c
&\equiv &
\bar s_\alpha \gamma^\mu (1-\gamma_5) c_\alpha 
\, 
\bar c_\beta \gamma_\mu (1-\gamma_5) d_\beta
,
\label{eq:q2c}
\\
Q_3 
&\equiv &
\bar s_\alpha \gamma^\mu (1-\gamma_5) d_\alpha 
\, 
\sum_q^{N_f} \bar q_\beta \gamma_\mu (1-\gamma_5) q_\beta
,
\label{eq:q3}
\\
Q_4
&\equiv &
\bar s_\alpha \gamma^\mu (1-\gamma_5) d_\beta 
\, 
\sum_q^{N_f} \bar q_\beta \gamma_\mu (1-\gamma_5) q_\alpha
,
\label{eq:q4}
\\
Q_5
&\equiv &
\bar s_\alpha \gamma^\mu (1-\gamma_5) d_\alpha 
\, 
\sum_q^{N_f} \bar q_\beta \gamma_\mu (1+\gamma_5) q_\beta
,
\label{eq:q5}
\\
Q_6 
&\equiv &
\bar s_\alpha \gamma^\mu (1-\gamma_5) d_\beta 
\, 
\sum_q^{N_f} \bar q_\beta \gamma_\mu (1+\gamma_5) q_\alpha
,
\label{eq:q6}
\end{eqnarray}
where $\alpha$ and $\beta$ are the fundamental color indices, and the summation over $N_f$ goes up to the allowed flavors at the given scale.
The Hamiltonian of Eq. (\ref{eq:effhamimw}) keeps the same form down to $\mu = m_c$, but the Wilson coefficients run in the change of the scale.

The running is calculated in the next-to-leading order logarithmic approximation (NLLA) \cite{Buras:1991jm,Buchalla:1995vs,Yamanaka:2015ncb}.
Below $\mu = m_c$, the charm quark is integrated out.
The resulting $|\Delta S| =1$ effective Hamiltonian
becomes
\begin{equation}
{\cal H}_{eff} (\mu)
=
\frac{G_F}{\sqrt{2}} V_{us}^* V_{ud}
\sum_{i=1}^6
 z_i (\mu) Q_i (\mu)
 + {\rm H.c.} 
.
\label{eq:effhamibelowmc}
\end{equation}
Here we quote the values of Refs. \cite{Yamanaka:2015ncb,Yamanaka:2016fjj}:
\begin{equation}
{\bf z} (\mu = 1 \, {\rm GeV})
=
\left(
\begin{array}{c}
-0.107 \cr
1.02 \cr
1.76 \times 10^{-5} \cr
-1.39 \times 10^{-2}  \cr
6.37 \times 10^{-3} \cr
-3.45 \times 10^{-3} \cr
\end{array}
\right)
.
\end{equation}
We see that the Wilson coefficient of $Q_2$ is the largest.
This is because $Q_2$ is the sole tree level operator at $\mu =m_W$, and the others were radiatively generated.
Here we point that the coefficient of $Q_1$ is also important since the contribution of $Q_2$ obtains a factor of $1/N_c$ after the Fierz rearrangement of the color (see below).
The operators $Q_i$ ($i=3,\cdots 6$) cannot be neglected either, because they generate the $\phi$ meson which is impossible with $Q_1$ and $Q_2$.
We also note that $Q_5$ and $Q_6$, after Fierz transformation, couple to the chiral condensate which may enhance the overall effect (see below).

For the calculation of the crossing symmetric contribution, it is convenient to Fierz transform the $|\Delta S|=1$ four-quark operators $Q_i$ ($i=1,\cdots 6$).
The Fierz transform of Eqs. (\ref{eq:q1}), (\ref{eq:q2}), (\ref{eq:q3}), (\ref{eq:q4}), (\ref{eq:q5}), and (\ref{eq:q6}) are 
\begin{eqnarray}
Q_1
&=&
\frac{1}{3}
\bar s \gamma^\mu (1-\gamma_5) u \, \bar u \gamma_\mu (1-\gamma_5) d
+2
\sum_a \bar s \gamma^\mu (1-\gamma_5) t_a u \, \bar u \gamma_\mu (1-\gamma_5) t_a d 
\nonumber\\
&=&
\bar s \gamma_\mu (1-\gamma_5) d \, \bar u \gamma^\mu (1-\gamma_5) u
,
\label{eq:q1fierz}
\\
Q_2
&=&
\frac{1}{3} \bar s \gamma_\mu (1-\gamma_5) d \, \bar u \gamma^\mu (1-\gamma_5) u
+ 2 \sum_{a=1}^8 \bar s \gamma_\mu (1-\gamma_5) t_a d \, \bar q \gamma^\mu (1-\gamma_5) t_a q
,
\label{eq:q2fierz}
\\
Q_3
&=&
\frac{1}{3} \sum_{q=u,d,s} \bar s \gamma_\mu (1-\gamma_5) q \, \bar q \gamma^\mu (1-\gamma_5) d 
+ 2 \hspace{-0.5em} \sum_{q=u,d,s} \sum_{a=1}^8 \bar s \gamma_\mu (1-\gamma_5) t_a q \, \bar q \gamma^\mu (1-\gamma_5) t_a d 
,
\ \ \ \ \ 
\label{eq:q3fierz}
\\
Q_4
&=&
\frac{1}{3}
\bar s \gamma^\mu (1-\gamma_5) d \, \sum_{q=u,d,s} \bar q \gamma_\mu (1-\gamma_5) q 
+2
\sum_a \bar s \gamma^\mu (1-\gamma_5) t_a d \, \sum_{q=u,d,s} \bar q \gamma_\mu (1-\gamma_5) t_a q 
\nonumber\\
&=&
\sum_{q=u,d,s} \bar s \gamma_\mu (1-\gamma_5) q \, \bar q \gamma^\mu (1-\gamma_5) d 
,
\label{eq:q4fierz}
\\
Q_5
&=&
-\frac{2}{3} \sum_{q=u,d,s} \bar s (1+\gamma_5) q \, \bar q (1-\gamma_5) d
- 4 \sum_{q=u,d,s} \sum_a \bar s (1+\gamma_5) t_a q \, \bar q (1-\gamma_5) t_a d 
,
\label{eq:q5fierz}
\\
Q_6
&=&
\frac{1}{3}
\bar s \gamma^\mu (1-\gamma_5) d \sum_{q=u,d,s} \bar q \gamma_\mu (1+\gamma_5) q
+2
\sum_a \bar s \gamma^\mu (1-\gamma_5) t_a d \sum_{q=u,d,s} \bar q \gamma_\mu (1+\gamma_5) t_a q
\nonumber\\
&=&
-2 \sum_{q=u,d,s} \bar s (1+\gamma_5) q \, \bar q (1-\gamma_5) d 
,
\label{eq:q6fierz}
\end{eqnarray}
where $t_a$ is the generator of the color $SU(3)_c$ group.
The summation over the fundamental color indices runs inside each Dirac bilinear, so the indices ($\alpha$ and $\beta$) have been omitted.
As for Eqs. (\ref{eq:q1fierz}), (\ref{eq:q4fierz}), and (\ref{eq:q6fierz}), we also displayed in the first equalities the Fierz rearrangement of the fundamental color indices to form color singlet Dirac bilinears.
We note that an additional minus sign contributes due to the anticommutation of fermion operators.
This sign change is important since there may be interference with crossing symmetric graphs.

\begin{figure}[htb]
 \begin{center}
\includegraphics[width=18cm]{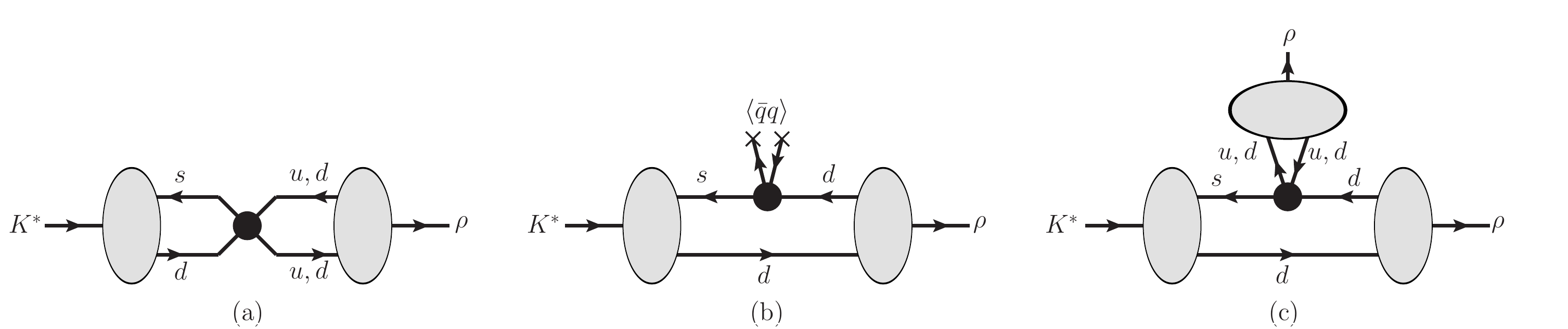}
\caption{ 
Factorization of the $|\Delta S|=1$ vector meson vertices ($|\Delta S|=1$ meson transition), with (a) the two-quark process, (b) the one-quark process, and (c) three-meson interaction.
The double crosses with "$\langle \bar q q \rangle$" denote the chiral condensate $\langle 0 | \bar q q |0 \rangle$ ($q=d,s$).
The black blob denotes the $|\Delta S| =1$ four-quark interaction.
There are similar diagrams with the $\rho$ meson replaced by $\omega$ and $\phi$ mesons.
}
\label{fig:factorization}
 \end{center}
\end{figure}

We use the standard factorization to derive the $|\Delta S| =1$ vector meson interaction from the $|\Delta S| =1$ four-quark interaction of Eq. (\ref{eq:effhamibelowmc}). 
We first construct the $|\Delta S| =1$ meson transition in the factorization with vacuum saturation approximation \cite{Lee:1972px,Shrock:1978dm}.
It works as
\begin{eqnarray}
\langle \rho | \bar s \gamma^\mu d \, \bar q \gamma_\mu q | K^* \rangle
&\approx&
\langle 0 | \bar s \gamma^\mu d | K^* \rangle \langle \rho | \bar q \gamma_\mu q | 0 \rangle
,
\label{eq:vacuumsaturation1}
\\
\langle \rho | \bar s d \, \bar d d | K^* \rangle
&\approx&
\langle \rho | \bar s d | K^* \rangle \langle 0 | \bar d d | 0 \rangle
,
\label{eq:vacuumsaturation2}
\end{eqnarray}
where $q=u,d$.
We note that the vacuum saturation approximation gives the leading contribution in the large $N_c$ expansion in the mesonic sector.
The $|\Delta S| =1$ four-quark interaction has two distinct contributions, as shown in 
Figs. \ref{fig:factorization}(a) and \ref{fig:factorization}(b). 
The first contribution (a) is the factorization into two meson tadpoles [see Eq. (\ref{eq:vacuumsaturation1})].
It requires the decay constants of vector mesons, as 
\begin{eqnarray}
\langle 0 | \bar u \gamma_\mu u | \rho \rangle
&=&
\frac{1}{\sqrt{2}} \varepsilon_\mu m_\rho f_\rho
,
\\
\langle 0 | \bar d \gamma_\mu d | \rho \rangle
&=&
-\frac{1}{\sqrt{2}}\varepsilon_\mu m_\rho f_\rho
,
\\
\langle 0 | \bar q \gamma_\mu q | \omega \rangle
&=&
\frac{1}{\sqrt{2}}\varepsilon_\mu m_\omega f_\omega
 \ \ \ (q=u,d)
,
\\
\langle 0 | \bar s \gamma_\mu s | \phi \rangle
&=&
\varepsilon_\mu m_\phi f_\phi
,
\end{eqnarray}
where $\varepsilon_\mu$ is the polarization of the vector meson, and $m_\rho = 770$ MeV, $ f_\rho = 216$ MeV, $m_\omega = 783$ MeV, $ f_\omega = 197$ MeV, $m_\phi = 1020$ MeV, and $ f_\phi = 233$ MeV \cite{Neubert:1997uc,Jansen:2009hr,Grossmann:2015lea,Straub:2015ica,Chang:2018aut,Sun:2018cdr}.
The second contribution [Fig. \ref{fig:factorization} (b)] is the factorization into scalar matrix elements [see Eq. (\ref{eq:vacuumsaturation2})].
It appears from the Fierz transformation of $Q_5$ and $Q_6$.
The chiral condensates relevant in this regard are
$\langle 0 | \bar ss  | 0 \rangle \approx \langle 0 | \bar dd | 0 \rangle \approx -\frac{m_\pi^2 f_\pi^2}{m_u+m_d} \approx -(269\, {\rm MeV})^3$ \cite{McNeile:2012xh}.
They are obtained at the appropriate renormalization scale $\mu = 1$ GeV with $m_u \approx 2.7$ MeV and $m_d \approx 5.9$ MeV \cite{Tanabashi:2018oca}, calculated in the two-loop level renormalization group evolution \cite{Tarasov:1980au,Gorishnii:1983zi}.
The scalar matrix element of the vector meson is derived by using the result of the calculation of the  chiral extrapolation of the vector meson mass in lattice QCD \cite{Leinweber:2001ac,Rios:2008zr,Bavontaweepanya:2018yds,Guo:2018zvl,Molina:2020xxx}.
As derived in Appendix~\ref{sec:scalar_matrix_elements},
we obtain
\begin{align}
 B_{\rho K^\ast} &\equiv \langle \rho^0 |\bar{s}d|K^{\ast 0}\rangle =-1.14 \,\, \text{GeV}, \\
 B_{\omega K^\ast} &\equiv \langle \omega |\bar{s}d|K^{\ast 0}\rangle=1.88 \,\, \text{GeV}, \\ 
 B_{\phi K^\ast} &\equiv \langle \phi |\bar{d}s|K^{\ast 0}\rangle=2.14 \,\, \text{GeV}.
\end{align}

By using the above parameters, 
the lagrangian of the weak vector meson transition is given by
\begin{eqnarray}
{\cal L}_{V K^*}
&=&
V_{ud}V_{us}^*
\hspace{-0.5em}
\sum_{V = \rho , \omega , \phi}
\hspace{-0.5em}
g_{ V  K^*} {V}^\nu K^*_\nu
+{\rm H.c.}
,
\label{eq:meson-transition}
\end{eqnarray}
where $\rho^\nu$, $\omega^\nu$ and $\phi^\nu$ are the field operators of the 
$\rho^0$, $\omega$, and $\phi$ mesons, respectively.
The coupling constants are given by
\begin{eqnarray}
g_{ \rho  K^*}
&=&
\frac{G_F}{\sqrt{2}}
\biggl[
\biggl(
z_1 +\frac{1}{3}z_2  
-\frac{1}{3}z_3 - z_4 
\biggr)
m_{K^*} f_{K^*}
m_\rho \frac{f_\rho}{\sqrt{2}}
-
\biggl(
\frac{2}{3} z_5 + 2 z_6
\biggr)
B_{\rho K^\ast} 
\langle 0 | \bar s s+\bar dd | 0 \rangle
\biggr]
\nonumber\\
&=&
4.4 \times 10^{-8} {\rm GeV}^2 
,
\\
g_{ \omega  K^*}
&=&
\frac{G_F}{\sqrt{2}}
\biggl[
\biggl(
z_1 +\frac{1}{3}z_2 
+\frac{7}{3}z_3 +\frac{5}{3}z_4
+2 z_5 + \frac{2}{3} z_6
\biggr)
m_{K^*} f_{K^*}
m_\omega \frac{f_\omega}{\sqrt{2}}
-\biggl(
\frac{2}{3} z_5 + 2 z_6
\biggr)
B_{\omega K^\ast} 
\langle 0 | \bar s s+\bar dd | 0 \rangle
\biggr]
\nonumber\\
&=&
3.4 \times 10^{-8} {\rm GeV}^2
,
\\
g_{ \phi  K^*}
&=&
\frac{G_F}{\sqrt{2}}
\biggl[
\biggl(
\frac{4}{3}z_3 +\frac{4}{3}z_4  
+z_5 + \frac{1}{3} z_6  
\biggr)
m_{K^*} f_{K^*}
m_\phi f_\phi
-\biggl(
\frac{2}{3} z_5 + 2 z_6
\biggr)
B_{\phi K^\ast} 
\langle 0 | \bar s s+\bar dd | 0 \rangle
\biggr]
\nonumber\\
&=&
-6.6 \times 10^{-9} {\rm GeV}^2
.
\end{eqnarray}

Let us also construct the weak three-meson interactions.
Again by using the vacuum saturation approximation, we have
\begin{eqnarray}
\langle \rho \, | \bar q \gamma_\mu q\, \bar s \gamma^\mu d | K^* \rho \rangle
&\approx&
\langle 0 | \bar q \gamma_\mu q | \rho \rangle \langle \rho | \bar s \gamma^\mu d | K^* \rangle
,
\ \ \ \ 
\label{eq:vacuumsaturation3}
\end{eqnarray}
with $q=u,d$.
The weak three-vector meson interaction is then
\begin{eqnarray}
&&
{\cal L}^V_{V' K^*}
=
V_{ud} V_{us}^*
\hspace{-1em}
\sum_{V,V'=\rho , \omega , \phi}
\hspace{-1em}
g^V_{ V'  K^*} V_\mu {V'}^\nu i \overleftrightarrow \partial^\mu K^*_\nu
+{\rm H.c.}
, \ \ \ 
\label{eq:three-meson}
\end{eqnarray}
where $A \overleftrightarrow \partial^\mu B \equiv A (\partial^\mu B) - (\partial^\mu A) B$.
The coupling constants are given by
\begin{eqnarray}
g^\rho_{ V'  K^*}
&=&
\frac{G_F}{\sqrt{2}}
\biggl[
z_1 +\frac{1}{3} z_2
-\frac{1}{3}z_3 - z_4
\biggr]
m_{\rho} \frac{f_{\rho}}{\sqrt{2}} c_{V^\prime K^\ast}
\nonumber\\
&=&
( 2.4 \times 10^{-7} ) \times c_{V^\prime K^\ast}
,
\\
g^\omega_{ V'  K^*}
&=&
\frac{G_F}{\sqrt{2}}
\biggl[
z_1 +\frac{1}{3} z_2
+\frac{7}{3}z_3 + \frac{5}{3} z_4
+2 z_5 +\frac{2}{3} z_6
\biggr]
m_{\omega} \frac{f_{\omega}}{\sqrt{2}} c_{V^\prime K^\ast}
\nonumber\\
&=&
( 2.0 \times 10^{-7} ) \times c_{V^\prime K^\ast}
,
\\
g^\phi_{ V'  K^*}
&=&
\frac{G_F}{\sqrt{2}}
\biggl[
\frac{4}{3} z_3 +\frac{4}{3} z_4
+z_5 + \frac{1}{3} z_6
\biggr]
m_{\phi} f_{\phi} c_{V^\prime K^\ast}
\nonumber\\
&=&
( -2.6 \times 10^{-8} )\times c_{V^\prime K^\ast}
,
\label{eq:three-meson-coupling}
\end{eqnarray}
where $V' = \rho , \omega , \phi$. 
The coefficients $c_{V^\prime K^\ast}$ are obtained as
the relative strength of the meson transition ${\rm Tr}[V^\mu T^\dagger V_\mu + V^{\mu\dagger} T V^\dagger_\mu]$,
where $T$ is the SU(3) ladder operator given by the Gell-Mann matrices 
$\lambda_a$ $(a=1,...,8)$ as
$T=\frac{1}{2\sqrt{2}}(\lambda_6+i\lambda_7)$.
As a result, we obtain $c_{\rho K^\ast}=1/\sqrt{2}$, $c_{\omega K^\ast}=-1/\sqrt{2}$ and $c_{\phi K^\ast} = -1$.
For the amplitudes of the weak three-vector meson interaction,
we used the approximate relation $\langle \rho (p') | \bar s \gamma^\mu d | K^* (p) \rangle \approx  -(p^\mu + p'^\mu)\varepsilon^{(\rho)\nu} \varepsilon^{(K^*)*}_\nu$.

\subsection{One-loop level calculation of the EDM of charged leptons}
In this subsection, we 
perform the one-loop level calculation of the lepton EDM which is given by the amplitudes 
shown in Fig.~\ref{fig:electron_EDM_long-distance}.
The diagrams in 
Figs.~\ref{fig:electron_EDM_long-distance}(a) and \ref{fig:electron_EDM_long-distance}(a$^\prime$) 
are the contribution with the weak interaction of three vector mesons, while the diagrams in Figs.~\ref{fig:electron_EDM_long-distance} (b) and (b$^\prime$) are that with the strong interaction.

The scattering amplitudes with the weak three-meson interaction 
in Figs.~\ref{fig:electron_EDM_long-distance} (a) and (a$^\prime$)
are given by
\begin{align}
 i{\cal M}^{K^\ast}_{(a)}
 =&
 - ieJm_{K^\ast}f_{K^\ast}I_{dsll}\left(\frac{em^2_\rho}{g_\gamma}\right)^2
 \sum_{V,V^\prime=\rho,\omega,\phi}
\frac{c_Vc_{V^\prime} }{q^2-m^2_{V}}
\notag\\
& \times  
 \int\frac{d^4 k}{(2\pi)^4}
 \frac{\bar{u}_{l}(p-q)\gamma_\mu[(\pslash-\kslash)+m_l]
 \left[g^{V}_{V^\prime K^\ast}(2k-q)\cdot\varepsilon\gamma^\mu +g^{V^\prime}_{V K^\ast}(k+q)^\mu\epslash\right]\gamma_5u_{l}(p)}{(k-q)^2[(p-k)^2-m^2_{l}] [(k-q)^2-m^2_{V^\prime}][k^2-m^2_{K^\ast}]}, \label{eq:amp_a_Kast} \\
 i{\cal M}^{K^\ast}_{(a^\prime)}=& 
+ ieJm_{K^\ast}f_{K^\ast}I_{dsll}\left(\frac{em^2_\rho}{g_\gamma}\right)^2
 \sum_{V,V^\prime=\rho,\omega,\phi}
\frac{c_Vc_{V^\prime} }{q^2-m^2_{V}}
\notag\\
& \times  
 \int\frac{d^4 k}{(2\pi)^4}
 \frac{\bar{u}_{l}(p-q)\gamma_\mu[(\pslash-\kslash)-m_l]
 \left[g^{V}_{V^\prime K^\ast}(2k-q)\cdot\varepsilon\gamma^\mu +g^{V^\prime}_{V K^\ast}(k-2q)^\mu\epslash\right]\gamma_5u_{l}(p)}{k^2[(p-k)^2-m^2_{l}] [k^2-m^2_{V^\prime}][(k-q)^2-m^2_{K^\ast}]} . \label{eq:amp_aprime_Kast}
\end{align}
The masses of leptons ($l=e,\mu,\tau$) are given by 
$m_e = 0.510 998 950$ MeV,
$m_\mu = 0.105658$ GeV, and $m_\tau = 1.77686$ GeV~\cite{Tanabashi:2018oca}.
The coefficients $c_V, c_{V^\prime}$ are $c_\rho=1$, $c_\omega=1/3$, $c_\phi=-\sqrt{2}/3$.
In the soft photon limit ($q^2 \sim 0$, $p\cdot q\sim 0$), the denominators of the integrands
in Eqs.~\eqref{eq:amp_a_Kast}-\eqref{eq:amp_aprime_Kast} are rewritten as
\begin{align}
&\frac{1}{(k-q)^2[(p-k)^2-m^2_{l}] [(k-q)^2-m^2_{V^\prime}][k^2-m^2_{K^\ast}]}
=\Gamma(4) \int_{0}^{1}dz_1\int_{0}^{z_1}dz_2\int_{0}^{z_2}dz_3\frac{1}{[\ell^2_a-\Delta_a]^4}, \\
&\frac{1}{k^2[(p-k)^2-m^2_{l}] [k^2-m^2_{V^\prime}][(k-q)^2-m^2_{K^\ast}]}
=\Gamma(4) \int_{0}^{1}dz_1\int_{0}^{z_1}dz_2\int_{0}^{z_2}dz_3\frac{1}{[\ell^2_{a^\prime}-\Delta_a]^4},
\end{align}
where 
\begin{align}
 \ell^\mu_a&=k^\mu-z_3p^\mu-(z_1-z_3)q^\mu,  \\
 \ell^\mu_{a^\prime}&=k^\mu-z_3p^\mu-(1- 
 z_1
 )q^\mu , \\
 \Delta_a&=
 m^2_{K^\ast}+(m^2_{V^\prime}-m^2_{K^\ast})z_1-m^2_{V^\prime}z_2+m^2_{l}z^2_3 .
\end{align}
The numerators of the integrands in Eqs.~\eqref{eq:amp_a_Kast}-\eqref{eq:amp_aprime_Kast} are reduced to 
\begin{align}
 &\bar{u}_l(p-q)\gamma_\mu [(\pslash-\kslash)+
 m_l ]
 \left[g^{V}_{V^\prime K^\ast}(2k-q)\cdot\varepsilon\gamma^\mu 
 +g^{V^\prime}_{V K^\ast}(k+q)_\mu \epslash
 \right]\gamma_5u_{l}(p)  \notag\\
 &=\bar{u}_l(p-q)
 \left[ 4 g^{V}_{V^\prime K^\ast} m_l z_3 (3-2z_1+z_3) p\cdot\varepsilon
+ g^{V^\prime}_{V K^\ast} m_l z_3 \left(2 p\cdot\varepsilon +\qslash\epslash\right)
 \right]
 \gamma_5u_{l}(p) + \cdots \\
& \bar{u}_{l}(p-q)\gamma_\mu[(\pslash-\kslash)-m_l]
 \left[g^{V}_{V^\prime K^\ast}(2k-q)\cdot\varepsilon\gamma^\mu +g^{V^\prime}_{V K^\ast}(k-2q)^\mu\epslash\right]\gamma_5u_{l}(p) \notag\\
 &=  
 -\bar{u}_{l}(p-q) \left[
 4g^{V}_{V^\prime K^\ast}  m_l z_3(3-2z_1+z_3)p\cdot\varepsilon
 +g^{V^\prime}_{V K^\ast}  m_l (3z_3-2z_1-2)\qslash\epslash
\right] \gamma_5u_{l}(p) + \cdots
\end{align}
where the terms which do not contribute to the EDM are suppressed. 
By performing the integrals with respect to $\ell_a$ and $\ell_{a^\prime}$ 
for Eqs.~\eqref{eq:amp_a_Kast}-\eqref{eq:amp_aprime_Kast},
the amplitudes for Eqs.~\eqref{eq:amp_a_Kast}-\eqref{eq:amp_aprime_Kast}
are reduced to 
\begin{align}
 i{\cal M}^{K^\ast}_{(a)}=&  -  \frac{i}{(4\pi)^2}em_l Jm_{K^\ast}f_{K^\ast}I_{dsll}\left(\frac{em^2_\rho}{g_\gamma}\right)^2
 \sum_{V,V^\prime=\rho,\omega,\phi}\frac{c_V c_{V^\prime}}{m^2_V}\bar{u}_{l}(p-q) \notag\\
 &\times \left[
 \int_{0}^{1}dz_1\int_{0}^{z_1}dz_2\int_{0}^{z_2}dz_3 
 \frac{2g^{V}_{V^\prime K^\ast}  
 z_3(3-2z_1+z_3)
 +2g^{V^\prime}_{V K^\ast} 
 z_3
 }{\Delta^2_a}\right]
 \sigma^{\mu\nu}q_\nu\varepsilon_\mu\gamma_5 u_{l}(p) , \label{eq:amp_a_final} \\
 i{\cal M}^{K^\ast}_{(a^\prime)}=&  -  \frac{i}{(4\pi)^2}em_l Jm_{K^\ast}f_{K^\ast}I_{dsll}\left(\frac{em^2_\rho}{g_\gamma}\right)^2
 \sum_{V,V^\prime=\rho,\omega,\phi}\frac{c_V c_{V^\prime}}{m^2_V}\bar{u}_{l}(p-q) \notag\\
 &\times \left[
 \int_{0}^{1}dz_1\int_{0}^{z_1}dz_2\int_{0}^{z_2}dz_3 
 \frac{2g^{V}_{V^\prime K^\ast} 
 z_3(3-2z_1+z_3)
 +g^{V^\prime}_{V K^\ast} 
 (3z_3-2z_1-2)
 }{\Delta^2_a}\right]
 \sigma^{\mu\nu}q_\nu\varepsilon_\mu\gamma_5 u_{l}(p) , \label{eq:amp_aprime_final}
\end{align}
respectively, where we use the Gordon identity
\begin{align}
 \bar{u}_l(p-q)\left[(2p-q)^\mu-i\sigma^{\mu\nu}q_\nu\right]\gamma_5u_l(p)=0 .
\end{align}
The integrals in Eqs.~\eqref{eq:amp_a_final} and \eqref{eq:amp_aprime_final},  
\begin{align} 
 I^{(a)}_1 &=
 \int_{0}^{1}dz_1\int_{0}^{z_1}dz_2\int_{0}^{z_2}dz_3\frac{z_3(3-2z_1+z_3)}{\Delta^2_a} \label{eq:I(a)_1} \\
 I^{(a)}_2 &=
 \int_{0}^{1}dz_1\int_{0}^{z_1}dz_2\int_{0}^{z_2}dz_3\frac{z_3}{\Delta^2_a} \label{eq:I(a)_2} \\
 I^{(a^\prime)}_2 &=
 \int_{0}^{1}dz_1\int_{0}^{z_1}dz_2\int_{0}^{z_2}dz_3
 \frac{3z_3-2z_1-2}{\Delta^2_a} 
 \label{eq:I(aprime)_2}
\end{align}
are performed numerically, with the results summarized in Table~\ref{table:Numerical_integral_a}.
For $I^{(a)}_2$, the analytic form is obtained as shown in Appendix~\ref{sec:Ia2}.

 \begin{table}[htbp]
  \caption{\label{table:Numerical_integral_a} Numerical values of the integrals in Eqs.~\eqref{eq:I(a)_1}-\eqref{eq:I(aprime)_2} for the leptons
  $l=e,\mu,\tau$ and the vector mesons $V^\prime=\rho^0,\omega,\phi$,
given in units of GeV$^{-4}$.}
 \begin{center}
    \begin{tabular}{c|ccc||c|ccc||c|ccc}  
     \hline
     $I^{(a)}_1$&$\rho^0$ &$\omega$ &$\phi$
     &$I^{(a)}_{2}$ &$\rho^0$ &$\omega$ &$\phi$
     &$I^{(a^\prime)}_{2}$ &$\rho^0$ &$\omega$ &$\phi$
      \\ \hline
     $e$& 23.3&22.6 &13.6 
		& $e$& 14.1& 13.6& 8.20    
		& $e$& -82.1& -79.6& -47.6    
		 \\
     $\mu$& 4.82 & 4.70& 3.03  
		 & $\mu$& 2.97& 2.89& 1.85 
		 & $\mu$& -25.9& -25.2& -15.5 
				 \\
     $\tau$& 0.209& 0.206& 0.153  
		 & $\tau$& 0.137& 0.134& 0.0972  
		 & $\tau$& -4.68& -4.56& -2.96 
				  \\ \hline
    \end{tabular}
 \end{center}
 \end{table}

The amplitudes $i{\cal M}^{K^\ast}_{(a)}$ and $i{\cal M}^{K^\ast}_{(a^\prime)}$ are for the contributions with the $K^\ast$ propagator.
In addition, the amplitudes with the $\bar{K}^\ast$ propagator, denoted by $i{\cal M}^{\bar{K}^\ast}_{(a)}$ and $i{\cal M}^{\bar{K}^\ast}_{(a^\prime)}$, 
also contribute to the EDM.
If we restrict to the CP violation, we have
$i{\cal M}^{\bar{K}^\ast}_{(a)} = i{\cal M}^{K^\ast}_{(a)}$ and $i{\cal M}^{\bar{K}^\ast}_{(a^\prime)} = i{\cal M}^{K^\ast}_{(a^\prime)}$.
Thus the total scattering amplitude with the weak three-vector meson interactions is given by
\begin{align}
 i{\cal M}^{\rm SM}_{(a)} & =  i{\cal M}_{(a)} +  i{\cal M}_{(a^\prime)}, \label{eq:amp_MaSM}
\end{align}
where
\begin{align}
 i{\cal M}_{(a)}
&=i{\cal M}^{K^\ast}_{(a)} + i{\cal M}^{\bar{K}^\ast}_{(a)} 
 = 2i{\cal M}^{K^\ast}_{(a)} , \label{eq:amp_Ma} \\
 i{\cal M}_{(a^\prime)}
&=i{\cal M}^{K^\ast}_{(a^\prime)} + i{\cal M}^{\bar{K}^\ast}_{(a^\prime)} 
 = 2i{\cal M}^{K^\ast}_{(a^\prime)}. \label{eq:amp_Maprime}
\end{align}

In a similar manner, 
the charged lepton EDM contributions with the strong three-vector meson interactions 
shown in 
Figs.~\ref{fig:electron_EDM_long-distance}$(b)$ and \ref{fig:electron_EDM_long-distance}($b^\prime$)
are also calculated.
The scattering amplitudes of the diagrams $(b)$ and $(b^\prime)$ are obtained as
\begin{align}
 i{\cal M}^{K^\ast}_{(b)} = i{\cal M}^{\bar{K}^\ast}_{(b)}
 =&  -  ieJm_{K^\ast}f_{K^\ast}I_{dsll}\left(\frac{em^2_\rho}{g_\gamma}\right)^2
 \sum_{V,V^\prime=\rho^0,\omega,\phi}
\frac{c_Vc_{V^\prime} }{q^2-m^2_{V}} g_V g_{V^\prime K^\ast}
\notag\\
& \times  
 \int\frac{d^4 k}{(2\pi)^4}
 \frac{\bar{u}_{l}(p-q)\gamma_\mu[(\pslash-\kslash)+m_l]
 \left[
 (2k-q)\cdot\varepsilon \gamma^\mu +\varepsilon^\mu (2\qslash-\kslash) 
 -(k+q)^\mu\epslash
 \right]\gamma_5u_{l}(p)}{(k-q)^2[(p-k)^2-m^2_{l}] [(k-q)^2-m^2_{V^\prime}]
 [(k-q)^2-m^2_{K^\ast}][k^2-m^2_{K^\ast}]}, \label{eq:amp_b_Kast} \\
  i{\cal M}^{K^\ast}_{(b^\prime)} = i{\cal M}^{\bar{K}^\ast}_{(b^\prime)} =& 
 ieJm_{K^\ast}f_{K^\ast}I_{dsll}\left(\frac{em^2_\rho}{g_\gamma}\right)^2
 \sum_{V,V^\prime=\rho^0,\omega,\phi}
\frac{c_Vc_{V^\prime} }{q^2-m^2_{V}} g_V g_{V^\prime K^\ast}
\notag\\
& \times  
 \int\frac{d^4 k}{(2\pi)^4}
 \frac{\bar{u}_{l}(p-q)\gamma_\mu[(\pslash-\kslash)-m_l]
 \left[
 (2k-q)\cdot\varepsilon \gamma^\mu +\varepsilon^\mu (2\qslash-\kslash) 
 -(k+q)^\mu\epslash
 \right]\gamma_5u_{l}(p)}{k^2[(p-k)^2-m^2_{l}] [k^2-m^2_{V^\prime}]
 [(k-q)^2-m^2_{K^\ast}][k^2-m^2_{K^\ast}]} .
 \label{eq:amp_bprime_Kast}
\end{align}
The coupling constant $g_V$ ($V=\rho^0, \omega, \phi$) is defined as 
$\sqrt{2}g_\rho = -\sqrt{2}g_\omega = g_\phi = g$. 
In the soft photon limit, 
the denominators of the integrands
in Eqs.~\eqref{eq:amp_b_Kast}-\eqref{eq:amp_bprime_Kast} are rewritten as
\begin{align}
&\frac{1}{(k-q)^2[(p-k)^2-m^2_{l}] [(k-q)^2-m^2_{V^\prime}][(k-q)^2-m^2_{K^\ast}][k^2-m^2_{K^\ast}]}
=\Gamma(5) \int_{0}^{1}dz_1\int_{0}^{z_1}dz_2\int_{0}^{z_2}dz_3\int_{0}^{z_3}dz_4
 \frac{1}{[\ell^2_b-\Delta_b]^5}, \\
&\frac{1}{k^2[(p-k)^2-m^2_{l}] [k^2-m^2_{V^\prime}][(k-q)^2-m^2_{K^\ast}][k^2-m^2_{K^\ast}]}
=\Gamma(5) \int_{0}^{1}dz_1\int_{0}^{z_1}dz_2\int_{0}^{z_2}dz_3\int_{0}^{z_3}dz_4
 \frac{1}{[\ell^2_{b^\prime}-\Delta_b]^5},
\end{align}
where 
\begin{align}
 \ell^\mu_b&=k^\mu-z_4p^\mu-(z_1-z_4)q^\mu,  \\
 \ell^\mu_{b^\prime}&=k^\mu-z_4p^\mu-(z_1-z_2)q^\mu, \\
 \Delta_b&=
 m^2_{K^\ast}+(m^2_{V^\prime}-m^2_{K^\ast})z_2-m^2_{V^\prime}z_3+m^2_{l}z^2_4 .
\end{align}
Performing the integrals with respect to $\ell_b$ and $\ell_{b^\prime}$,
Eqs.~\eqref{eq:amp_b_Kast}-\eqref{eq:amp_bprime_Kast} are reduced to
\begin{align}
 i{\cal M}^{K^\ast}_{(b)}=&  -  \frac{  4  i}{(4\pi)^2}eJ 
 m_l
m_{K^\ast}f_{K^\ast}I_{dsll}\left(\frac{em^2_\rho}{g_\gamma}\right)^2
 \sum_{V,V^\prime=\rho^0,\omega,\phi}\frac{c_V c_{V^\prime}}{m^2_V} g_Vg_{V^\prime K^\ast}
 \bar{u}_{l}(p-q) \notag\\
 &\times \left[
 \int_{0}^{1}dz_1\int_{0}^{z_1}dz_2\int_{0}^{z_2}dz_3\int_{0}^{z_3}dz_4
 \frac{
 (1-z_1)(1-2z_4)+1-z_4^2
 }{\Delta^3_b}\right]
 \sigma^{\mu\nu}q_\nu\varepsilon_\mu\gamma_5 u_{l}(p) , \label{eq:amp_b_final} \\
 i{\cal M}^{K^\ast}_{(b^\prime)}=&  -  \frac{ 4 i}{(4\pi)^2}eJ 
m_l m_{K^\ast}f_{K^\ast}I_{dsll}\left(\frac{em^2_\rho}{g_\gamma}\right)^2
 \sum_{V,V^\prime=\rho^0,\omega,\phi}\frac{c_V c_{V^\prime}}{m^2_V} g_Vg_{V^\prime K^\ast}
 \bar{u}_{l}(p-q) \notag\\
 &\times \left[
 \int_{0}^{1}dz_1\int_{0}^{z_1}dz_2\int_{0}^{z_2}dz_3\int_{0}^{z_3}dz_4
 \frac{
 (z_1-z_2)(1-  2 z_4) + 1-z_4^2
 }{ 
\Delta^3_b
}\right]
 \sigma^{\mu\nu}q_\nu\varepsilon_\mu\gamma_5 u_{l}(p) . \label{eq:amp_bprime_final}
\end{align}
The numerical results of the integrals
\begin{align}
 I^{(b)} &=
 \int_{0}^{1}dz_1\int_{0}^{z_1}dz_2\int_{0}^{z_2}dz_3\int_{0}^{z_3}dz_4
 \frac{
 (1-z_1)(1-2z_4)+1-z_4^2
 }{\Delta^3_b} , \label{eq:I(b)} \\
 I^{(b^\prime)} & =
 \int_{0}^{1}dz_1\int_{0}^{z_1}dz_2\int_{0}^{z_2}dz_3\int_{0}^{z_3}dz_4
 \frac{ 
 (z_1-z_2)(1-  2 z_4) + 1-z_4^2
 }{\Delta^3_b} \label{eq:I(bprime)}
,
\end{align}
are summarized in Table~\ref{table:Numerical_integral_b}.

 \begin{table}[htbp]
  \caption{\label{table:Numerical_integral_b} Numerical values of the integrals in Eqs.~\eqref{eq:I(b)}-\eqref{eq:I(bprime)} for the leptons
  $l=e,\mu,\tau$ and the vector mesons $V^\prime=\rho^0,\omega,\phi$,
  given in units of GeV$^{-6}$.}
 \begin{center}
    \begin{tabular}{c|ccc||c|ccc}
     \hline
     $I^{(b)}$ &$\rho^0$ &$\omega$ &$\phi$
     &$I^{(b^\prime)}$ &$\rho^0$ &$\omega$ &$\phi$
      \\ \hline
     $e$& 13.7&13.3 &7.93  
		 &  $e$& 13.7&13.3 &7.93  
		 \\
        $\mu$& 4.32 & 4.19& 2.56 
 &     $\mu$& 4.32 & 4.19& 2.56 
				 \\
     $\tau$& 0.714& 0.697& 0.447   
		 &     $\tau$& 0.714& 0.697& 0.447
				  \\ \hline
    \end{tabular}
 \end{center}
 \end{table}

Finally, the total scattering amplitude with the strong three-vector meson interactions is obtained as
\begin{align}
 i{\cal M}^{\rm SM}_{(b)} & = i{\cal M}_{(b)} + i{\cal M}_{(b^\prime)}, \label{eq:amp_MbSM} \\
 i{\cal M}_{(b)}
&=i{\cal M}^{K^\ast}_{(b)} + i{\cal M}^{\bar{K}^\ast}_{(b)} 
 = 2i{\cal M}^{K^\ast}_{(b)} , \label{eq:amp_Mb} \\
 i{\cal M}_{(b^\prime)}
&=i{\cal M}^{K^\ast}_{(b^\prime)} + i{\cal M}^{\bar{K}^\ast}_{(b^\prime)} 
 = 2i{\cal M}^{K^\ast}_{(b^\prime)} . \label{eq:amp_Mbprime}
\end{align}

\section{\label{sec:analysis}Results and analysis}

\subsection{\label{sec:result}Numerical results}

From the scattering amplitudes derived in the previous section,
we obtain the hadronic long distance contributions to the EDMs of charged leptons.
From the amplitudes $ i{\cal M}_{(a)}$ and $ i{\cal M}_{(a^\prime)}$ of Eqs.~\eqref{eq:amp_Ma} and \eqref{eq:amp_Maprime}, 
we obtain the EDMs generated by the weak three-vector meson interactions
as 
\begin{align}
 d^{\rm SM}_{(a)e}    &=d_{(a)e}+d_{(a^\prime)e}  
=
  3.67 \times 10^{-40 } 
 \, e\, {\rm cm}, \label{eq:dSM_ae} \\
 d^{\rm SM}_{(a)\mu}&=d_{(a)\mu}+d_{(a^\prime)\mu} 
 =
-1.04 \times 10^{-40 } 
 \, e\, {\rm cm}, \label{eq:dSM_amu} \\
 d^{\rm SM}_{(a)\tau}&=d_{(a)\tau}+d_{(a^\prime)\tau} 
 =
 - 1.12 \times 10^{-37}  
 \, e\, {\rm cm}. \label{eq:dSM_atau} 
\end{align}
Similarly, 
the amplitudes $i{\cal M}_{(b)}$ and $i{\cal M}_{(b^\prime)}$ of Eqs.~\eqref{eq:amp_Ma} and \eqref{eq:amp_Maprime}
give the contribution from the strong three-vector meson interaction:
\begin{align}
 d^{\rm SM}_{(b)e}    &
= 
2.13 \times 10^{-40} 
 \, e\, {\rm cm}, \label{eq:dSM_be}\\
 d^{\rm SM}_{(b)\mu}&
=
 1.39 \times 10 ^{-38} 
 \, e\, {\rm cm}, \label{eq:dSM_bmu}\\
 d^{\rm SM}_{(b)\tau}&
=
  3.89 \times 10^{-38} 
 \, e\, {\rm cm}. \label{eq:dSM_btau} 
\end{align}

We finally obtain the EDMs of $e,\mu$, and $\tau$ generated by the hadronic long distance contributions
as
\begin{eqnarray}
d_e^{\rm SM} 
&=& d^{\rm SM}_{(a)e} + d^{\rm SM}_{(b)e} =
 5.80 \times 10^{-40} e \, {\rm cm}
,
\\
d_\mu^{\rm SM} 
&=& d^{\rm SM}_{(a)\mu} + d^{\rm SM}_{(b)\mu} =
 1.38 \times 10^{-38} 
e \, {\rm cm}
,
\\
d_\tau^{\rm SM} 
&=& d^{\rm SM}_{(a)\tau} + d^{\rm SM}_{(b)\tau} =
 -7.32 \times 10^{-38} 
e \, {\rm cm}
.
\end{eqnarray}

These values are much larger than the estimation at the four-loop level (\ref{eq:eEDMquark}), (\ref{eq:muEDMquark}), and (\ref{eq:tauEDMquark}).
The most important reason of this enhancement is due to the relevance of the typical hadronic momenta in the loop.
We recall that the elementary (quark) level contribution only had a typical momentum of $O(m_W) \sim O(m_t)$, and this feature, together with the GIM mechanism, forced the EDM of charged leptons to have a suppression factor $m_b^2 m_c^2 m_s^2$ due to the cancellation between terms with very close values.
We note that the GIM mechanism is also working at the hadron level.
However, the cancellation among contributions with different flavors becomes much milder thanks to the fact that the typical momentum is replaced by the mass of the heaviest hadrons of each diagram.
This is probably a general property of the hadronic CP violation in the SM, as similar enhancement is also relevant for the case of the EDM of neutron \cite{Czarnecki:1997bu,McKellar:1987tf,Seng:2014lea,Pospelov:2013sca} or nuclei \cite{Yamanaka:2015ncb}.
In this sense, the fact that the elementary contribution to the EDM appears only at the four-loop level is not truly essential in this strong suppression, but rather the GIM mechanism (or the antisymmetry of the Jarlskog invariant) is the main cause \cite{Pospelov:2013sca}.

We should also comment on the observable effect of the electron EDM in experiments.
The EDM of the electron is usually measured through the paramagnetic atomic or molecular systems, since the relativistic effect enhances its effect \cite{Carrico:1968zz,sandars1,sandars2,Flambaum:1976vg,Sandars:1975zz,Labzovskii,Sushkovmolecule,kelly,Kozlov:1994zz,Kozlov:1995xz,flambaumfr1,nayak1,nayak3,nayak2,Natarajrubidium,Mukherjeefrancium,Dzuba:2009mw,Nataraj:2010vn,flambaumybftho,Porsev:2012zx,Roberts:2013zra,Chubukov:2014rba,abe,sunaga,Radziute:2015apa,Denis,Skripnikov,Sunaga:2018lja,Sunaga:2018pjn,Malika:2019jhn,Fazil:2019esp}.
However, these systems also receive contribution from other CP violating mechanisms such as the CP-odd electron-nucleon interaction or the nuclear Schiff moment.
Previously, the EDM of charged lepton was believed to be extremely small and the CP-odd electron-nucleon interaction was thought to be the dominant effect, with a benchmark value equivalent to the electron EDM as $d_e^{(eN)}\sim (10^{-39}-10^{-37} ) e$ cm for paramagnetic systems 
\cite{Pospelov:2013sca,Yamanaka:2015ncb,Yamanaka:2017mef,PhysRevA.93.062503} .
By considering the strong enhancement at the hadronic level, we just obtained a value of the electron EDM which lies in this range.
It is then an interesting question to quantify which one, between the electron EDM and the CP-odd electron-nucleon interaction, gives the leading contribution at the atomic level.

\subsection{\label{sec:error}Error bar analysis}

We now assess the uncertainty of our calculation.
The first important source of systematics is the nonperturbative effect of the renormalization of the $|\Delta S|=1$ four-quark operators.
This was quantified to be about 10\%, by looking at the variation of the Wilson coefficient of $Q_2$ in the NLLA in the range of the scale $\mu = 0.6$ GeV to $\mu = m_c = 1.27$ GeV \cite{Buras:1991jm,Buchalla:1995vs,Yamanaka:2015ncb}.
Another major systematics comes from the factorization of the vector meson matrix elements.
According to the large $N_c$ analysis, the vacuum saturation approximation should work up to $O(N_c^{-1})$ correction.
To be conservative, we set the error bar associated to it to 40\%.

\begin{figure}[t]
 \begin{center}
\includegraphics[width=18cm]{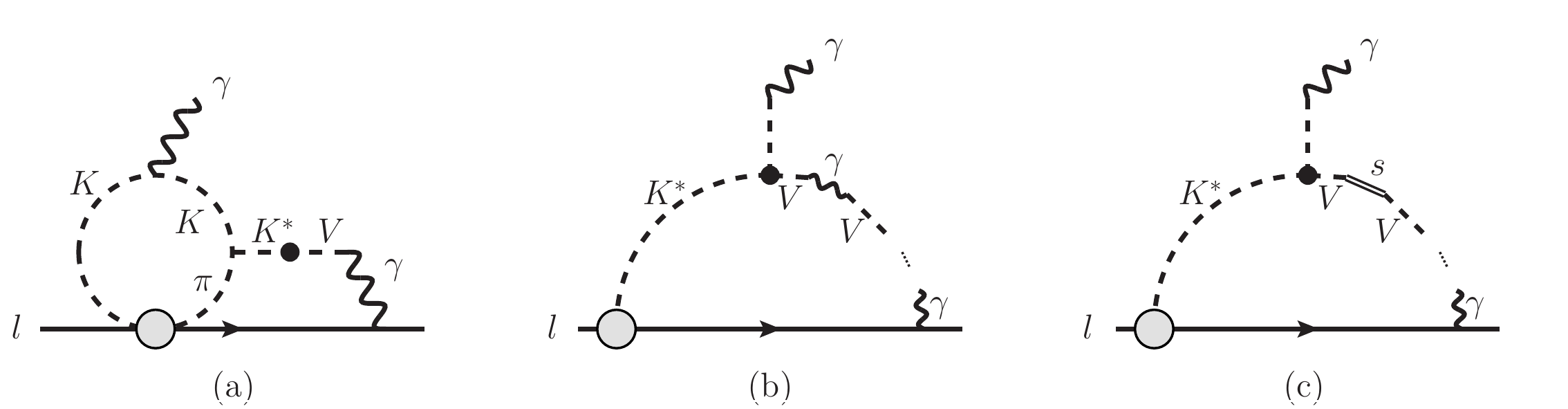}
\caption{ Higher order contributions to the EDM of charged leptons within the HLS. Here $V=\rho^0, \omega, \phi$ is the neutral vector meson, and $s$ is the would-be Nambu-Goldstone boson of the HLS.} 
\label{fig:HLS_higer}
 \end{center}
\end{figure}

Let us now see the uncertainty related to the use of the HLS.
The important point is that the one-loop level diagrams we evaluated are not divergent, so that the uncertainty due to the counterterms is absent and the stability of the coupling constants is guaranteed.
However, we have to comment on the contribution from higher order terms.
The first process to be noted is the two-loop level diagram with pseudoscalar mesons [see Fig.~\ref{fig:HLS_higer} (a)].
This contribution is the most straightforward higher order effect of the HLS.
The second type of higher order process is the mixing of vector mesons with photons [see Fig.~\ref{fig:HLS_higer} (b)].
In the mass eigenstate basis, it is possible to take into account this mixing to all orders, 
but if we restrict the analysis to the leading order, as done in this work, our particle basis and the mass one coincide.
The third contribution is the would-be Nambu-Goldstone mode which may appear 
as scalar propagator insertions
in between vector meson propagations due to the choice of Feynman gauge [see Fig.~\ref{fig:HLS_higer} (c)].
This is again a higher order effect in the HLS, just like the mixing between photons and vector mesons.
We also note that the explicit flavor SU(3) breaking effect is a higher order of the HLS, 
which has already partially been included in our calculations by introducing the physical meson masses and decay constants.
We also have to comment on the contribution from the 
other heavier hadrons which were overlooked in this paper.
Here we consider the axial vector meson $K_1$(1270) which, in the viewpoint of the mass difference, should be the most important hadron among the neglected ones, and show that its contribution is likely to be subleading.
First, the decay constant of $K_1$ is not particularly enhanced, $f_{K_1} \sim 170$ MeV \cite{Nardulli:2005fn}.
Regarding the other hadron matrix elements, the values do not exist in the literature, but it is possible to show that they are not enhanced either.
The axial vector matrix element $\langle \rho | \bar d \gamma_\mu \gamma_5 s | K_1 \rangle$ corresponds to the quark spin, so there should be a suppression due to the destructive interference generated by successive gluon emissions/absorptions \cite{Yamanaka:2013zoa,Yamanaka:2014lva}.
The pseudoscalar matrix element $\langle \rho | \bar d \gamma_5 s | K_1 \rangle$ has also no reasons to be enhanced, since this receives contribution from the Nambu-Goldstone boson pole, which is suppressed by the $K$ meson mass in the present case.
We can consider that the effect of $K_1$(1270) and other heavier hadrons is part of the higher order contribution of the HLS.
We associate 
the theoretical uncertainty coming from the entire higher order process mentioned above with the expansion parameter, estimated to be $\sim 50\%$~\cite{Harada:1992np}. 
In all, we conclude that the theoretical uncertainty is $ 70\%$.

A potentially interesting way to quantify the hadronic contribution to the electron EDM is to calculate the hadronic three-point correlator on lattice and then attach it to 
a lepton line to form the EDM amplitude.
This approach is actually used to quantify the hadronic light-by-light scattering contribution to the muon anomalous magnetic moment \cite{Blum:2019ugy}. 
Of course the calculation will not be easy since the three-point correlator must have a $|\Delta S|=1$ four-quark operator in the intermediate state, but this kind of analysis was already done in the context of $K \to \pi \pi$ decay \cite{Bai:2015nea}, so it should not be impossible.

\section{Conclusion}

\begin{figure}[htb]
 \begin{center}
\includegraphics[width=0.7\linewidth,clip]{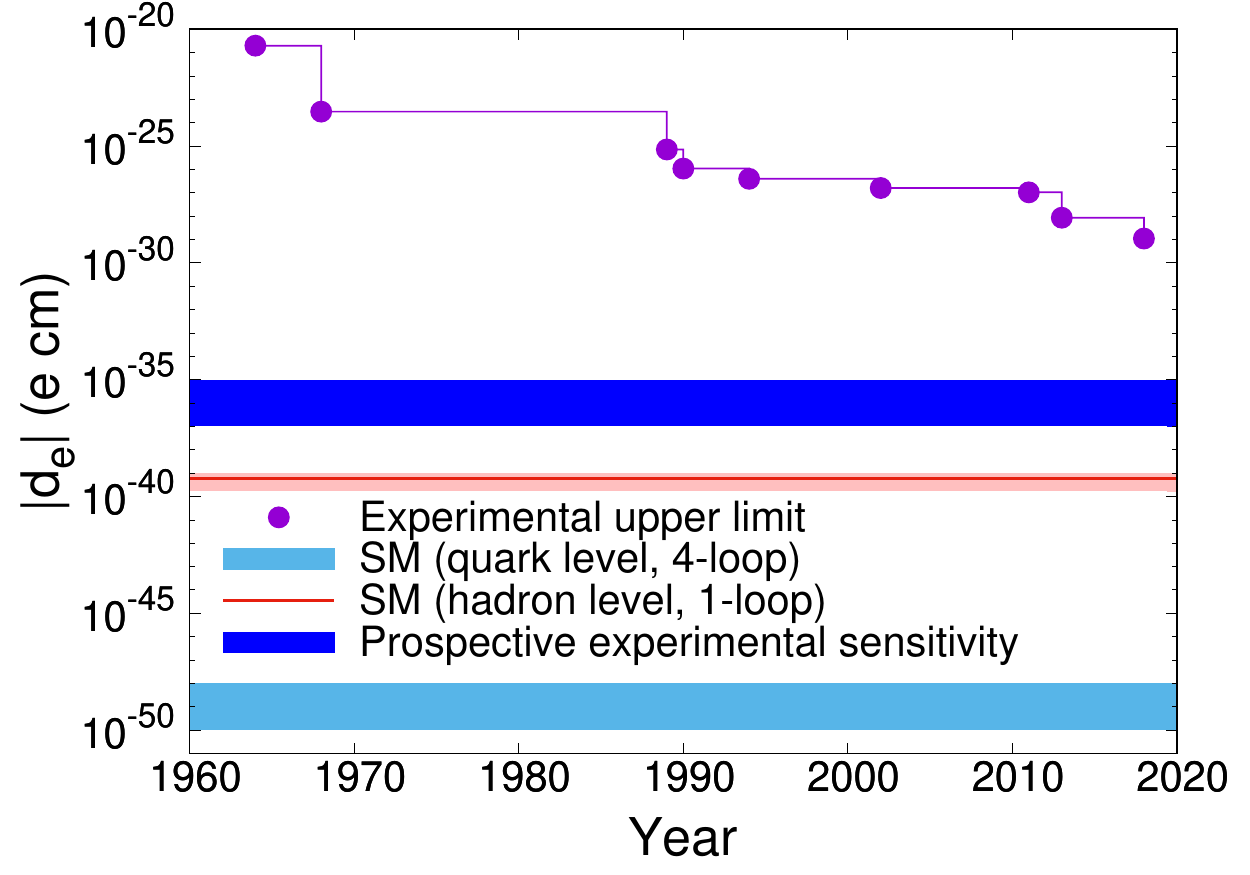}
\caption{
Plot of the SM predictions of the electron EDM compared with the history of the records of the experimental upper limits \cite{Sandars:1964zz,Weisskopf:1968zz,Player:1970zz,Murthy:1989zz,Abdullah:1990nh,Commins:1994gv,Regan:2002ta,Hudson:2011zz,Baron:2013eja,Andreev:2018ayy}.
The pink band is denoting the uncertainty of the hadronic contribution.
}
\label{fig:electronEDM_record}
 \end{center}
\end{figure}

In conclusion, we evaluated for the first time the hadron level contribution to the EDM of charged leptons in the SM, where the CP violation is generated by the physical complex phase of the CKM matrix.
As a result, we found that this long distance effect is much larger than the previously known one, which was estimated at the elementary level.
We could also rigorously show that, in the perturbative elementary level calculation at all orders, the EDM of charged leptons is always suppressed by quark mass factors due to the GIM mechanism.
The main reason of the enhancement at the hadronic level is because 
we could avoid additional factors of $m_{b,c,s}^2/m_{W,t}^2$ by embedding the heavy $W$ boson or top quark contribution into the $|\Delta S|=1$ low energy constants while keeping loop momenta of $O(\Lambda_{\rm QCD})\sim 1$ GeV.
In Fig. \ref{fig:electronEDM_record}, we plot the EDM of the electron in the SM compared with the progress of the experimental accuracy.
The electron EDM obtained in this work is $d_e \sim 10^{-39}e$ cm, which is still well below the current sensitivity of molecular beam experiments.
The EDM experiments are however improving very fast, and we have to be very sensitive to their progress and to proposals with new ideas, with some of them claiming to be able to go beyond the level of $O(10^{-35})e$ cm in statistical sensitivity \cite{Vutha:2017pej}.

Our next object would be to extend this analysis to 
the flavor violation disagreeing with the SM, recently suggested by the measurements of the decay of $B$ mesons at several $B$ factory experiments \cite{Lees:2012xj,Lees:2013uzd,Huschle:2015rga,Sato:2016svk,Hirose:2016wfn,Hirose:2017dxl,Aaij:2015yra,Aaij:2017uff,Aaij:2017deq,Aaij:2014ora,Aaij:2015oid,Wehle:2016yoi,Aaij:2017vbb,Aaij:2019wad,Hiller:2003js,Crivellin:2019qnh,Ikeno:2019tkh}, and that of $K$ meson decay of KOTO experiment \cite{KOTOresult,Kitahara:2019lws}.
In the analysis of the conjunction of the EDM with the $B$ meson decay, we also have to include the effect of heavy quarks, which has been omitted in this work.

\begin{acknowledgments}
The authors thank T. Kugo for useful discussions.
They also thank T. Morozumi for pointing out the contribution of the semi-leptonic penguin diagram.
A part of the numerical computation in this work was carried out at the Yukawa Institute Computer Facility.
This  work is supported  in  part  by  the  Special  Postdoctoral
Researcher Program  (SPDR) of RIKEN (Y.Y.).
\end{acknowledgments}

\appendix
\section{Scalar matrix element of vector mesons}
\label{sec:scalar_matrix_elements}

In this appendix, the scalar matrix element of vector mesons, $B_V$, are derived. 
Up to the tree-level contribution, the scalar matrix elements of vector mesons 
are obtained from the vector-meson masses expanded in terms of the light quark mass as 
\begin{align}
   &m_\rho^2 =
 M^2 + B_\rho (m_u+m_d) + ... 
 = M^2 + b_1M^2B_0 (m_u+m_d) + ... ,
 \label{eq:rhomass} \\
  &m_\omega^2 = 
 M^2 + B_\omega (m_u+m_d) +  ... 
 = M^2 + (b_1+2b_3) M^2B_0(m_u+m_d) +  ... ,
 \label{eq:omegamass} \\
  &m_\phi^2 =
 M^2 + B_\phi (2m_s) + ... 
 = M^2 + (b_1+b_3)M^2B_0 (2m_s) + ... ,
  \label{eq:phimass}
\end{align}
where $M$, $b_1$, 
and $b_3$ are the low-energy parameters of the effective Lagrangians~\cite{Bavontaweepanya:2018yds,Guo:2018zvl}\footnote{We omit the term being proportional to the parameter $b_2$ in~\cite{Guo:2018zvl,Bavontaweepanya:2018yds}, which does not contribute to the scalar matrix element for the vector-meson transition because ${\rm tr}\left[\chi_{+} T\right]=0$.}
\begin{align}
 {\cal L}=\frac{b_1}{8}M^2{\rm tr}\left[F^{\mu\nu}F_{\mu\nu}\chi_{+}\right]
 +\frac{b_3}{8}M^2{\rm tr}\left[F_{\mu\nu}\right]{\rm tr}\left[F^{\mu\nu}\chi_+\right] . \label{eq:L_rhomass}
\end{align}
$B_0$ is given by
\begin{align}
  &B_0=-\frac{\langle 0 | \bar{u}u|0 \rangle}{f^2} = \frac{(269\,\text{MeV})^3}{(93\,\text{MeV})^2}\sim 2.25\,\text{GeV}.
\end{align}
The scalar matrix elements for the vector-meson transitions such as $\langle \rho |\bar{s}d|K^\ast\rangle$ are related to $B_\rho \approx \frac{1}{2} \langle \rho | \bar u u + \bar d d | \rho \rangle$ by replacing $\chi_+\sim 2B_0diag(m_u,m_d,m_s)$ in \eqref{eq:L_rhomass} with $\chi_+ T$, where $T$ is the SU(3) ladder operator.
In the flavor SU(3) limit, we obtain
\begin{align}
 \langle \rho^0 |\bar{s}d|K^{\ast 0}\rangle & \sim -\frac{1}{2} B_\rho = -\frac{1}{2} \left( b_1M^2B_0\right) , \\
 \langle \omega |\bar{s}d|K^{\ast 0}\rangle & \sim \frac{1}{2} B_\omega = \frac{1}{2} \left\{ (b_1+2b_3)M^2B_0 \right\} , \\
 \langle \phi |\bar{d}s|K^{\ast 0}\rangle & \sim \frac{1}{\sqrt{2}} B_\phi = \frac{1}{\sqrt{2}} \left\{ (b_1+b_3)M^2B_0\right\} .
\end{align}
The low-energy parameters $b_1$, $b_3$, $B_0$, and $M$ were determined by fitting the 
lattice QCD data in Ref.~\cite{Guo:2018zvl}, where three fitting results denoted by 
Fit 1, Fit 2, and Fit 3 
were obtained as summarized in Table~\ref{table:b_constants}. 
The scalar matrix elements obtained by using these parameters are also summarized in Table~\ref{table:b_constants}. 
In this study, we employ the averaged values obtained as $\langle \rho^0 |\bar{s}d|K^{\ast 0}\rangle=-1.14$ GeV, $\langle \omega |\bar{s}d|K^{\ast 0}\rangle=1.88$, GeV and $ \langle \phi |\bar{d}s|K^{\ast 0}\rangle=2.14$ GeV.

\begin{table}
 \begin{center}
  \caption{\label{table:b_constants} The low-energy parameters in Eqs.~\eqref{eq:rhomass}-\eqref{eq:phimass} obtained in Ref.~\cite{Guo:2018zvl}.}
  \begin{tabular}{lrrr}
   \hline\hline
   & Fit 1& Fit 2& Fit 3\\ \hline
   $M$ [MeV]& 759.3& 758.8& 757.0\\   
   $b_1$ [GeV$^{-2}$] & 1.2224& 1.3420& 1.4009\\ 
   $b_3$ [GeV$^{-2}$] & 0.5131& 0.3469& 0.4151\\
   $2B_0m/m^2_\pi$  & 1.141& 1.077 &1.106 \\ \hline
   $\langle \rho^0 |\bar{s}d|K^{\ast 0}\rangle$ [GeV]& $-1.09$& $-1.13$& $-1.21$\\ 
   $\langle \omega |\bar{s}d|K^{\ast 0}\rangle$ [GeV]&2.01 &1.72 &1.92 \\ 
   $\langle \phi |\bar{d}s|K^{\ast 0}\rangle$ [GeV]&2.19 &2.01 &2.21 \\ 
   \hline\hline
  \end{tabular}
 \end{center} 
\end{table}

\section{Analytic form of the Integral $I^{(a)}_2$}
\label{sec:Ia2}

The analytic forms of the integral $I^{(a)}_2$ in Eq.~\eqref{eq:I(a)_2} are summarized in this appendix.
For $m_{l}=m_e, m_\mu$, $I^{(a)}_2$ is obtained as 
\begin{align}
 I^{(a)}_2 = &\frac{1}{4 {m_{l}^4} \left( {m_{K^\ast}^2} - {m_{V^\prime}^2}\right)}   
 \left[
 {m_{V^\prime}}\,\sqrt{{m_{V^\prime}^2}-4\,{m_{l}^2}}\,
 \log \left({{ 
 ( {m_{V^\prime}^2} - 2\,{m_{l}^2} )
 - {m_{V^\prime}}\,\sqrt{{m_{V^\prime}^2}-4\,{m_{l}^2}} 
 }\over{2\,{m_{l}^2}}}\right) 
 \right. \notag \\
   &  \left.
   +{m_{K^\ast}}\,\sqrt{
 {m_{K^\ast}^2}-4\,{m_{l}^2}}\,
 \log \left({{
 {m_{K^\ast}} + \sqrt{{m_{K^\ast}^2}-4\, {m_{l}^2}}
 }\over{
 {m_{K^\ast}} - \sqrt{{m_{K^\ast}^2}-4\, {m_{l}^2}}
 }}\right) 
 +(m_{K^\ast}^2 - 2 m_{l}^2)\log\left(\frac{m_l^2}{m_{K^\ast}^2}\right)
 -(m_{V^\prime}^2 - 2 m_{l}^2)\log\left(\frac{m_l^2}{m_{V^\prime}^2}\right)  
\right] ,
\end{align}
which can be  expanded in terms of a small $m_{l}$ as 
\begin{align}
 I^{(a)}_2 = &
 \frac{1}{(m_{K^\ast}m_{V^\prime})^2}\left[
 \log\left(\frac{m_{K^\ast}}{m_{l}}\right) 
 - \frac{m_{K^\ast}^2}{m_{K^\ast}^2 - m_{V^\prime}^2} \log\left(\frac{m_{K^\ast}}{m_{V^\prime}}\right)
 - \frac{1}{4}
 \right]
 \notag\\
& 
 + \frac{2 m_{l}^2}{(m_{K^\ast}m_{V^\prime})^4} \left[
 (m_{K^\ast}^2 + m_{V^\prime}^2)\log\left(\frac{m_{K^\ast}}{m_{l}}\right)
 - \frac{m_{K^\ast}^4}{m_{K^\ast}^2 - m_{V^\prime}^2} \log\left(\frac{m_{K^\ast}}{m_{V^\prime}}\right)
 - \frac{26\, m_{K^\ast}^4 - 5\, m_{V^\prime}^4}{12(m_{K^\ast}^2 - m_{V^\prime}^2)} 
 \right] + \cdots .
\end{align}
On the other hand, $I^{(a)}_2$ for $m_{l} = m_\tau$ is given by the different form as 
\begin{align}
 I^{(a)}_2 = & \frac{1}{4m_\tau^4 (m_{K^\ast}^2 - m_{V^\prime}^2)} \left[
 2(2m_{\tau}^2 - m_{V^\prime}^2) \log\left(\frac{m_{K^\ast}}{m_{V^\prime}}\right)
 + 2(m_{V^\prime}^2 - m_{K^\ast}^2) \log\left(\frac{4m_{K^\ast}^2m_{\tau}^2 - m_{V^\prime}^4}{4m_{V^\prime}^2m_{\tau}^2 - m_{V^\prime}^4}\right)
 \right. \notag\\
 & +2\sqrt{4m_{V^\prime}^2m_{\tau}^2 - m_{V^\prime}^4}\left(
 \arctan\left(\frac{m_{V^\prime}^2}{
\sqrt{4m_{V^\prime}^2m_{\tau}^2 - m_{V^\prime}^4}
}\right)
 +\arctan\left(\frac{2m_{\tau}^2 - m_{V^\prime}^2}{
\sqrt{4m_{V^\prime}^2m_{\tau}^2 - m_{V^\prime}^4}
}\right)
 \right)
 \notag\\
 & +\left( m_{V^\prime}^2 - m_{K^\ast}^2 + i m_{K^\ast} \sqrt{4m_{\tau}^2 - m_{K^\ast}^2} \right)
 \log\left(2\frac{m_{K^\ast}^2-m_{V^\prime}^2}{m_{K^\ast}^2}
 \left(1-2\frac{4m_{V^\prime}^2m_{\tau}^2 - m_{V^\prime}^4}{4m_{K^\ast}^2m_{\tau}^2 - m_{V^\prime}^4}+i\frac{\sqrt{4m_{\tau}^2-m_{K^\ast}^2}}{m_{K^\ast}}  \right)
 \right)
 \notag\\
 & 
+\left( m_{V^\prime}^2 - m_{K^\ast}^2 - i m_{K^\ast} \sqrt{4m_{\tau}^2 - m_{K^\ast}^2} \right)
 \log\left(2\frac{m_{K^\ast}^2-m_{V^\prime}^2}{m_{K^\ast}^2}
 \left(1-2\frac{4m_{V^\prime}^2m_{\tau}^2 - m_{V^\prime}^4}{4m_{K^\ast}^2m_{\tau}^2 - m_{V^\prime}^4}-i\frac{\sqrt{4m_{\tau}^2-m_{K^\ast}^2}}{m_{K^\ast}}  \right)
 \right)
 \notag\\
 & -\left( m_{V^\prime}^2 - m_{K^\ast}^2 + i m_{K^\ast} \sqrt{4m_{\tau}^2 - m_{K^\ast}^2} \right)
 \log\left(2\frac{m_{K^\ast}^2-m_{V^\prime}^2}{m_{K^\ast}^2}
 \left(-1+2\frac{m_\tau^2}{m_{K^\ast}^2}\frac{
4
m_{K^\ast}^2
m_{\tau}^2 - m_{V^\prime}^4}{4
m_{V^\prime}^2 
m_{\tau}^2 - m_{V^\prime}^4}+i\frac{\sqrt{4m_{\tau}^2-m_{K^\ast}^2}}{m_{K^\ast}}  \right)
 \right)
 \notag\\
 & \left.
 -\left( m_{V^\prime}^2 - m_{K^\ast}^2 - i m_{K^\ast} \sqrt{4m_{\tau}^2 - m_{K^\ast}^2} \right)
 \log\left(2\frac{m_{K^\ast}^2-m_{V^\prime}^2}{m_{K^\ast}^2}
 \left(-1+2\frac{m_\tau^2}{m_{K^\ast}^2}\frac{4
m_{K^\ast}^2 
m_{\tau}^2 - m_{V^\prime}^4}{4
m_{V^\prime}^2
m_{\tau}^2 - m_{V^\prime}^4} - i\frac{\sqrt{4m_{\tau}^2-m_{K^\ast}^2}}{m_{K^\ast}}  \right)
 \right) 
  \right] .
\end{align}

\end{document}